\documentclass[preprint,showkeys]{revtex4}
\usepackage{graphicx}
\usepackage{amssymb}
\newcommand{\be}{\begin{eqnarray}}
\newcommand{\ee}{\end{eqnarray}}

\begin{document}
\title{Nucleon and flavor  form factors in a light front quark model in
AdS/QCD}
\author{Dipankar Chakrabarti and Chandan Mondal 
}                     
%
%
\affiliation{Department of Physics, Indian Institute of Technology Kanpur, Kanpur-208016, India.}
%
%

\begin{abstract}
Using the light front wave functions for the nucleons in a quark model in AdS/QCD, we calculate the nucleon electromagnetic form factors.
The flavor decompositions of the nucleon form factors are  calculated from the GPDs in this  model.
  We show that the nucleon form factors and their flavor  decompositions   calculated in AdS/QCD are in agreement with  experimental data.
%
\end{abstract} 
\maketitle

\section{ Introduction}

In recent years, AdS/QCD has emerged as one of the most promising techniques to unravel the structure of hadrons.   The Maldacena conjecture\cite{maldacena}   opened an attractive  channel to address a strongly coupled gauge theory in $d$ space-time dimensions by a dual weak coupling gravity theory in AdS$_{d+1}$ space. In the last decade, there have been several attempts to exploit this duality to resolve problems in QCD. 
The first application of AdS/CFT to QCD  was done by Polchinski and Strassler to address the hard scattering \cite{PS1} and in the context of  deep inelastic scattering(DIS)\cite{PS2}.  
To compare with the QCD, one needs to break the conformal invariance. There are two methods in the literature to achieve this goal, one is called hard wall model where a boundary is put in the AdS space where the wave functions are made to vanish and the other is called the soft wall model in which a confining potential is introduced in the AdS space which breaks the conformal invariance and generates the mass spectrum.

 The AdS/QCD for the baryon  has been developed by  several groups 
\cite{BT00,BT01,AC,SS,ads1,ads2}.
Though it gives only the semiclassical approximation of QCD, so far this method has  been successfully applied to  describe many hadron properties e.g., hadron mass spectrum, parton distribution functions, meson and nucleon form factors,  structure functions etc\cite{AC,BT1,BT1b,BT2,HSS}.   Recently it has been shown that the results with  the AdS/QCD wave functions remarkably agree with the experimental data for $\rho$ meson electroproduction \cite{forshaw}. Studies of the nucleon form factors with higher Fock  states have been done in \cite{hf1}.  The generalized parton distributions (GPDs)  are related to the  nucleon form factors by sum rules and thus are calculable in AdS/QCD. The GPDs  using the method developed in \cite{AC}  and also the charge and magnetization densities of the nucleons in the transverse plane have been studied  in \cite{vega} while the GPDs in a light front quark model in AdS/QCD have been studied in \cite{CM}.

In the understanding of the nucleon structures, the electromagnetic  form factors play very important roles. There are many experiments and theoretical investigations on this subject and  it remains to be a very active field of research for many years. We refer to the articles \cite{Gao,Hyd,Punjabi} for detailed review on this subject.  In\cite{AC}, the electric and magnetic form factors for the nucleons have been calculated in AdS/QCD and shown to  reasonably agree with the well known Kelly or Arrington fits.  In this work, we have calculated the Sachs form factors and also the Dirac and Pauli form factors for the nucleons and the flavor decompositions of them and compared with the experimental data.
 First we calculate the nucleon form factors $F_1$ and $F_2$ in a light front quark model with $SU(6$) spin-flavor symmetry using the light front wave functions obtained from AdS/QCD.  There are sum rules which relate the  nucleon form factors 
 to  the valence  GPDs $H^q_v(x,t)$ and $E^q_v(x,t)$  for up and down quarks, where $x$ is the longitudinal momentum fraction carried by the quark and $t=q^2$ is the square of the momentum transferred in the process.  The first moment of the GPDs   gives the flavor form factors $F_1^q$ and $F_2^q$  for the quarks.  
 The extraction of the experimental data for the flavors are tricky as it requires both proton and neutron form factor data at the same value of $Q^2$, which, in general, are not available. The   flavor decompositions of the experimental data on the nucleon form factors  have been done in \cite{Cates} and \cite{diehl13}. We use those data to compare our results. We show that 
 the results of nucleon form factors  as well as the  flavor form factors  obtained  in AdS/QCD are  in agreement with  the experimental data.  The main purpose of this paper is to analyze the flavor decompositions of the electromagnetic form factors. In this work we have evaluated   the flavor structures of the electromagnetic form factors of the nucleons and compared with the experimental results. This is the first time, that the flavor decompositions of the form factors have been studied in detail in  any AdS/QCD model.

 The paper is organized as follows. In Sec.\ref{ff}, we give a brief introduction about  electromagnetic  nucleon and flavor   form factors. The evaluation of the form factors in AdS/QCD has been discussed  in Sec.\ref{ads} and the results are compared with experimental data in Sec.\ref{comp}.  At the end, we provide a brief summary and conclusions in Sec.\ref{concl}.
  
 \section{ Form Factors}\label{ff}
 It is well known that the  matrix element of the electromagnetic current for nucleons requires two form factors namely Dirac and Pauli form factors:
 \be
 J_{had}^\mu=\bar{u}(p')\Big(\gamma^\mu F_1(q^2)+\frac{i\sigma^{\mu\nu}q_\nu}{2M}F_2(q^2)\Big)u(p),
 \ee
 where $q^2=(p'-p)^2=-2p'\cdot p+2M^2$ is square of the momentum transferred to the nucleon and $M$ is the nucleon mass.
 The normalizations of the form factors are given by
 $ F_1^p(0)=1, F_2^p(0)=\kappa_p=1.793$ for proton and $F_1^n(0)=0, F_2^n(0)=\kappa_n=-1.913$ for neutron. Cates et al.\cite{Cates} first decomposed the nucleon form factors into their flavor components. 
 Writing the hadronic current as the sum of quark currents we can  decompose the nucleon electromagnetic form factors into flavor dependent form factors.  Neglecting the strange  quark contribution, the hadronic matrix element for electromagnetic current can be  written as 
 \be
 J_{had}^\mu=\langle p/n\mid (e_u \bar{u}\gamma^\mu u+ e_d\bar{d}\gamma^\mu d)\mid p/n\rangle,
 \ee
 where $e_u$ and $e_d$ are the charges of $u$ and $d$ quarks in units of positron charge($e$).
Under  the charge  and isospin symmetry $\langle p\mid \bar{u}\gamma^\mu u\mid p\rangle= \langle n\mid \bar{d}\gamma^\mu d\mid n\rangle$, it is straightforward to write down the flavor decompositions of the nucleon form factors    as
\be
F_i^u=2F_i^p+F_i^n ~~{\rm and} ~~F_i^d=F_i^p+2F_i^n,~~(i=1,2),
\ee
with the normalizations $F_1^u(0)=2, F_2^u(0)=\kappa_u$ and $F_1^d(0)=1, F_2^d(0)=\kappa_d$ where the anomalous magnetic moments for the up and down quarks are $\kappa_u=2\kappa_p+\kappa_n=1.673$ and $\kappa_d=\kappa_p+2\kappa_n=-2.033$.
 It was shown in \cite{Cates} that though the ratio of Pauli and Dirac form factors for the proton $F_2^p/F_1^p \propto 1/Q^2$,  the $Q^2$ dependence is almost constant for the ratio of the quark form factors $F_2/F_1$ for both $u$ and $d$.  
 
There are two well established techniques to extract the nucleon form factors in  the experiments. One is from unpolarized scattering cross section data by Rosenbluth separation method. In this method, one extracts the Sachs  form factors which are expressed in terms of Dirac and Pauli form factors as
 \be
 G_E^{p/n}(Q^2)&=& F_1^{p/n}(Q^2)-\frac{Q^2}{4M^2}F_2^{p/n}(Q^2),\\
 G_M^{p/n}(Q^2)&=& F_1^{p/n}(Q^2)+F_2^{p/n}(Q^2),
 \ee
 where $Q^2=-q^2=-t$. The other method uses  either the target or the recoiled polarized proton along with the polarized lepton beam and is known as  polarization transfer technique in which  the ratios the Sachs form factors for the nucleons are measured. The relevant ratios for proton and neutron are defined as
 \be
  R^p=\frac{\mu_p G_E^p }{G_M^p}, ~~{\rm and }~~ R^n=\frac{\mu_n G_E^n }{G_M^n}.
  \ee
 The measurement of this ratio for proton was very  crucial as at large $Q^2$ the data from polarization method did not  match with the ratio obtained from the unpolarized cross section data by a Rosenbluth separation.  The Rosenbluth separation method was originally based on single photon exchange processes but it was shown that to resolve the discrepancy  one must include the two photon exchange amplitudes  in the method\cite{guichon} and thus paved the way for the multi-photon physics.
 
 We can also define the Sachs form factors for the quarks  in the same way as Dirac and Pauli form factors 
 \be
 G_{E,M}^p&=& e_u G_{E,M}^u+e_d G_{E,M}^d,\nonumber\\
 G_{E,M}^n&=& e_u G_{E,M}^d+e_d G_{E,M}^u,
 \ee
 i.e., $G_{E,M}^u=2 G_{E,M}^p+G_{E,M}^n$ and  $G_{E,M}^d=G_{E,M}^p+2 G_{E,M}^n$. Note that  the up quark contributions to the proton form factors are same as the down quark contributions to the neutron form factors as the charges of the quarks are factored out\cite{Qattan}. So, $G^q_{E/M}$ can be referred as the  Sachs form factors for the flavors.   Recently, there have been a lot of studies on flavor form factors. Qattan and Arrington \cite{Qattan} have analyzed the flavor decompositions of the form factors using a similar method as \cite{Cates} but included the two photon exchange processes in the Rosenbluth separation. In \cite{diehl13}, the experimental data for flavor form factors are used to fit the GPDs for up and down quarks  and also estimated the total angular momentum contribution of each flavor by evaluating Ji's sum rule. Following their convention, the flavor dependent contributions to the nucleon form factors are  referred in this article  as ``flavor form factors". In \cite{miller12}, the nucleon and flavor form factors have been studied in a light front quark-diquark model.
The flavor form factors are also discussed using a model for GPDs in \cite{harnandez}  and in a relativistic quark model based on Goldstone-boson exchange  in \cite{rohrmoser}.   In \cite{vega13},  a light front quark-diquark model has been derived in AdS/QCD and the Diarc and Pauli form factors for the quarks have been calculated.
The flavor form factors have also been studied in the SU(3) chiral quark-soliton model in \cite{silva}. 
\section{Nucleon and flavor form factors in AdS/QCD}\label{ads}
 For the derivation of the nucleon light front  wave functions in AdS/QCD we follow the works of 
 Brodsky and Teramond \cite{BT00,BT2}. For this work we consider only the soft wall model of AdS/QCD.
 The  action in the soft wall  model  is written as\cite{BT2}
\be
S&=&\int d^4x dz \sqrt{g}\Big( \frac{i}{2}\bar\Psi e^M_A\Gamma^AD_M\Psi -\frac{i}{2}(D_M\bar{\Psi})e^M_A\Gamma^A\Psi\nonumber\\
&&-\mu\bar{\Psi}\Psi-V(z)\bar{\Psi}\Psi\Big),
\ee
where $e^M_A=(z/R)\delta^M_A$ is the inverse  vielbein and $V(z)$ is the confining potential which breaks the conformal invariance and 
 $R$ is the AdS radius. 
  Discussions about the symmetry properties of the action and  bulk spinors can be found in \cite{Hong}  in the context of hard wall model and in \cite{hf1} in the context of soft wall model.
 The   Dirac equation in AdS derived from the above action is given by
\be
i\Big(z \eta^{MN}\Gamma_M\partial_N+\frac{d}{2}\Gamma_z\Big)\Psi -\mu R\Psi-RV(z)\Psi=0.\label{ads_DE}
\ee
With $z$ identified as the  light front transverse impact variable $\zeta$ which gives the separation of the  quark and 
gluonic constituents in the hadron, it is possible to extract  the light front wave functions for the hadron. 
In $d=4$ dimensions, $\Gamma_A=\{\gamma_\mu, -i\gamma_5\}$.
To map with the light front wave equation, we identify $z\to\zeta$, 
where $\zeta$
is the light front transverse variable,  and substitute $\Psi(x,\zeta)=e^{-iP\cdot x}\zeta^2\psi(\zeta)u(P)$ in Eq.(\ref{ads_DE})  and set $\mid \mu R\mid=\nu+1/2$ where  $\nu$ is related with the orbital angular momentum by $\nu=L+1$ .
 For linear confining potential  $U(\zeta)=(R/\zeta)V(\zeta)=\kappa^2\zeta$, we get the light front wave equation for the baryon in $2\times 2$ spinor representation as
 \be
 \big(-\frac{d^2}{d\zeta^2}-\frac{1-4\nu^2}{4\zeta^2}+\kappa^4\zeta^2&+&2(\nu+1)\kappa^2\Big)\psi_+(\zeta)\nonumber\\
 &=&{\cal{M}}^2\psi_+(\zeta),\\
 \big(-\frac{d^2}{d\zeta^2}-\frac{1-4(\nu+1)^2}{4\zeta^2}&+&\kappa^4\zeta^2+2\nu\kappa^2\Big)\psi_-(\zeta)\nonumber\\
 &=&{\cal{M}}^2\psi_-(\zeta).
 \ee 
  In case of mesons, the similar potential $\kappa^4\zeta^2$ appears in the Klein-Gordon equation  which can be generated by introducing a dilaton background $\phi=e^{\pm\kappa^2 z^2}$ in the AdS space which breaks the conformal invariance. But in case of baryon, the dilaton can be scaled out by a field redefinition\cite{BT2}. So, the confining potential for baryons cannot be produced by dilaton and is put in by hand in the soft wall model.
  The form of the confining  potential ($\kappa^4\zeta^2$) is unique for both the meson and baryon sectors \cite{BTD}.
The twist-3 nucleon wave functions in the soft wall model are obtained as  
\be
\psi_+(z) &= &\frac{\sqrt{2}\kappa^2}{R^2}z^{7/2} e^{-\kappa^2 z^2/2}, \label{psi+}\\
\psi_-(z) &= & \frac{\kappa^3}{R^2}z^{9/2} e^{-\kappa^2 z^2/2}. \label{psi-}
\ee
The spin non-flip amplitude for the electromagnetic transition in AdS space is related to the Dirac form factor in  physical space by\cite{BT2}
\be
\int && d^4x ~ dz \sqrt{g}{\bar \psi}_{p'}(x,z)e^A_M\Gamma_A   A^M(x,z)\psi_p(x,z)\nonumber\\
&& \sim  (2\pi)^4 \delta^4(p'-p-q)\epsilon_\mu{\bar u}(p')\gamma^\mu F_1(q^2)u(p),
\ee
where $A^M$ is an external electromagnetic field propagating in AdS space. With the holographic mapping of $z\to\zeta$ , the spin non-flip form factors are then given by
\be
F_{\pm}(Q^2)=g_{\pm} R^4\int \frac{dz}{z^4} V(Q^2,z)\mid\psi_{\pm}(z)\mid^2.\label{adsF1}
\ee
The coefficients $g_\pm$ are determined from the spin-flavor structure of the model.
The SU(6) spin-flavor symmetric quark model is constructed in the AdS/QCD by weighing the different Fock-state component by the charges and spin projections of the partons as dictated by the symmetry.  In the model, the probabilities to find a quark $q$ in proton or neutron with spin up or down are given by\cite{BT2}
$
N_{p\uparrow}^u=\frac{5}{3},~ N_{p\downarrow}^u=\frac{1}{3}, ~N_{p\uparrow}^d=\frac{1}{3}, N_{p\downarrow}^d=\frac{2}{3},~
N_{n\uparrow}^u=\frac{1}{3},~ N_{n\downarrow}^u=\frac{2}{3}, ~N_{n\uparrow}^d=\frac{5}{3}, N_{n\downarrow}^d=\frac{1}{3}.
$
The coefficients $g_\pm$ in Eq.(\ref{adsF1} ) for proton and neutron  are then 
$
g_p^+=N_{p\uparrow}^u e_u+N_{p\uparrow}^d e_d=1,~~ g_p^-=N_{p\downarrow}^u e_u+N_{p\downarrow}^d e_d=0,~
g_n^+=N_{n\uparrow}^u e_u+N_{n\uparrow}^d e_d=-\frac{1}{3},~~ g_n^-=N_{n\downarrow}^u e_u+N_{n\downarrow}^d e_d=\frac{1}{3}.
$
The Dirac form factors for the nucleons are thus obtained as
\be
F_1^p(Q^2)&=&R^4\int \frac{dz}{z^4} V(Q^2,z)\psi^2_+(z),\\
F_1^n(Q^2) &=& -\frac{1}{3}R^4\int \frac{dz}{z^4} V(Q^2,z)(\psi^2_+(z)-\psi^2_-(z)).
\ee
For Pauli form factors which involve nucleon spin flip,  the non-minimal coupling as proposed in \cite{AC} has to  be included. Then the formula for the Pauli form factor has been derived as \cite{BT2}
\be
F_2^{p/n}(Q^2) \sim \int \frac{dz}{z^3}\psi_+(z) V(Q^2,z)\psi_-(z).
\ee
The Pauli form factors are normalized to $F_2^{p/n}(0)=\kappa_{p/n}$ and using the wave functions (Eqs.(\ref{psi+}),(\ref{psi-})), the above formula can be rewritten as 
\be
F_2^{p/n}(Q^2) =\kappa_{p/n}R^4 \int \frac{dz}{z^4} V(Q^2,z) \psi^2_-(z).
\ee
Note that this  is consistent with the $SU(6)$ spin-flavor symmetry\cite{BT2,BT3}.
The bulk-to-boundary propagator for the  soft wall model
is given by
\be
 V(Q^2,z)=\Gamma(1+\frac{Q^2}{4\kappa^2})U\big(\frac{Q^2}{4\kappa^2},0,\kappa^2 z^2), \label{propagator}
 \ee
 where $U(a,b,z)$ is the Tricomi confluent hypergeometric function given by
 \be
 \Gamma(a)U(a,b,z)=\int_0^\infty e^{-zx} x^{a-1}(1+x)^{b-a-1}dx.
 \ee
We first calculate the form factors for the proton and compare with the available experimental data.  The only parameter in the theory $\kappa$ is fixed by fitting the ratio $Q^2F_2^p(Q^2)/F_1^p(Q^2)$ with the experimental data. All the other form factors and GPDs are calculated with this fixed value of $\kappa$.

\begin{figure}[htbp]
\small{(a)}\includegraphics[width=7.5cm,height=7.5cm,clip]{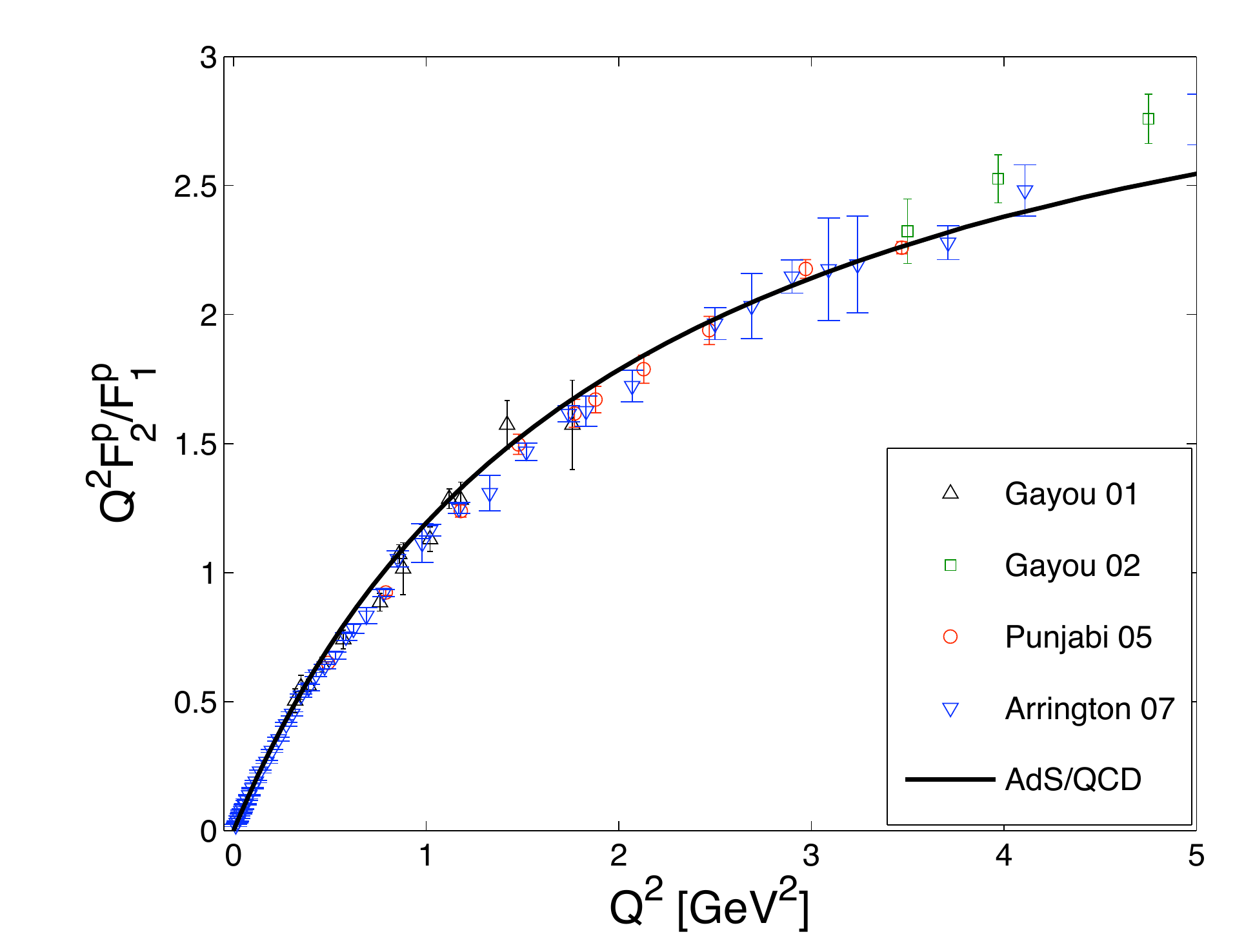}
\small{(b)}\includegraphics[width=7.5cm,height=7.5cm,clip]{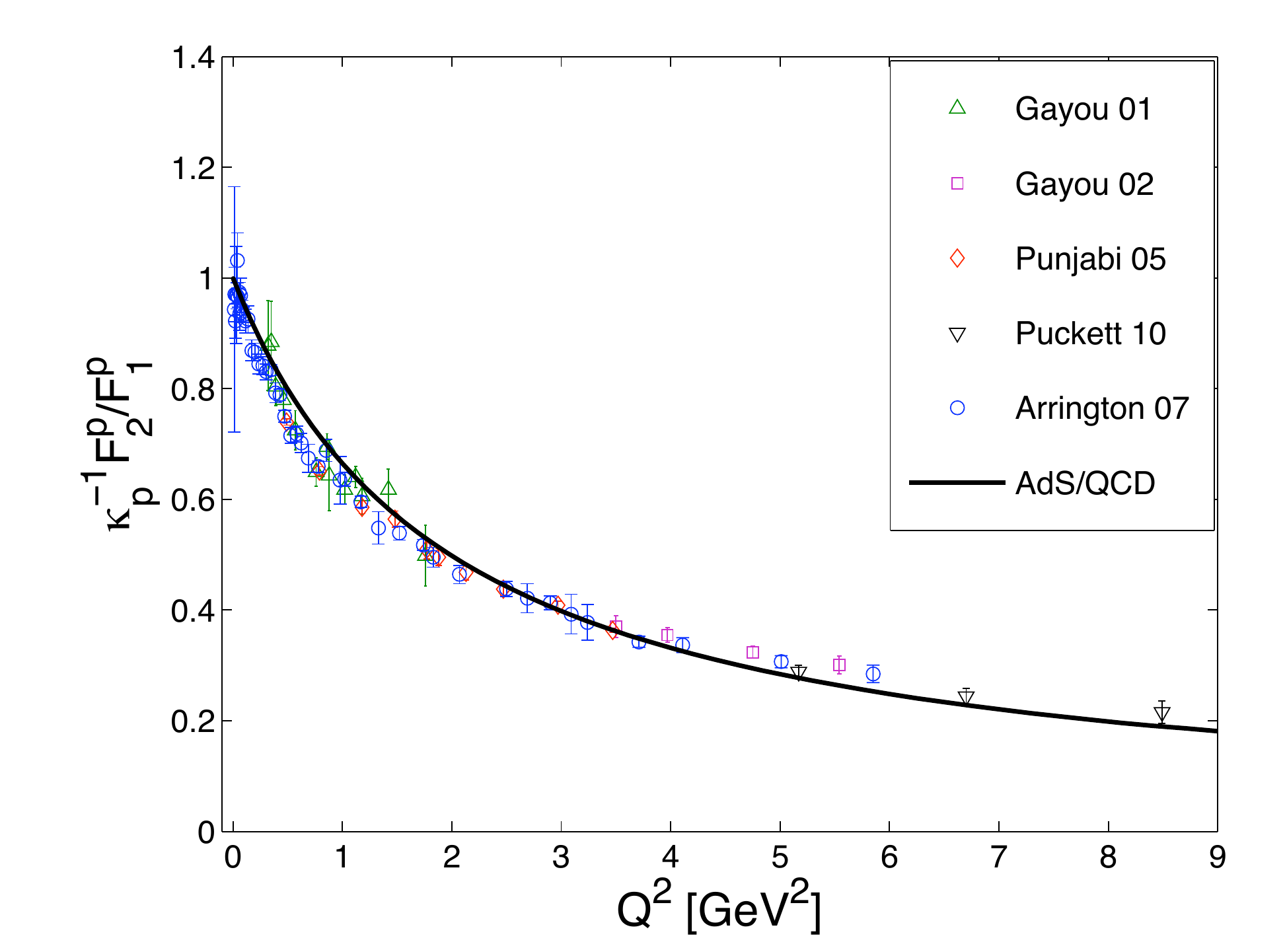}
\caption{\label{Fp_ratio}(Color online) The ratio of Pauli and Dirac form factors for the proton, (a) the ratio is multiplied by $Q^2=-q^2=-t$, (b) the ratio is divided by $\kappa_p$. The experimental  data are taken from Refs.  \cite{Gay1,Gay2,Arr,Pun,Puck}.}
\end{figure}
  \begin{figure}[htbp]
\small{(a)}\includegraphics[width=7.5cm,height=7.5cm,clip]{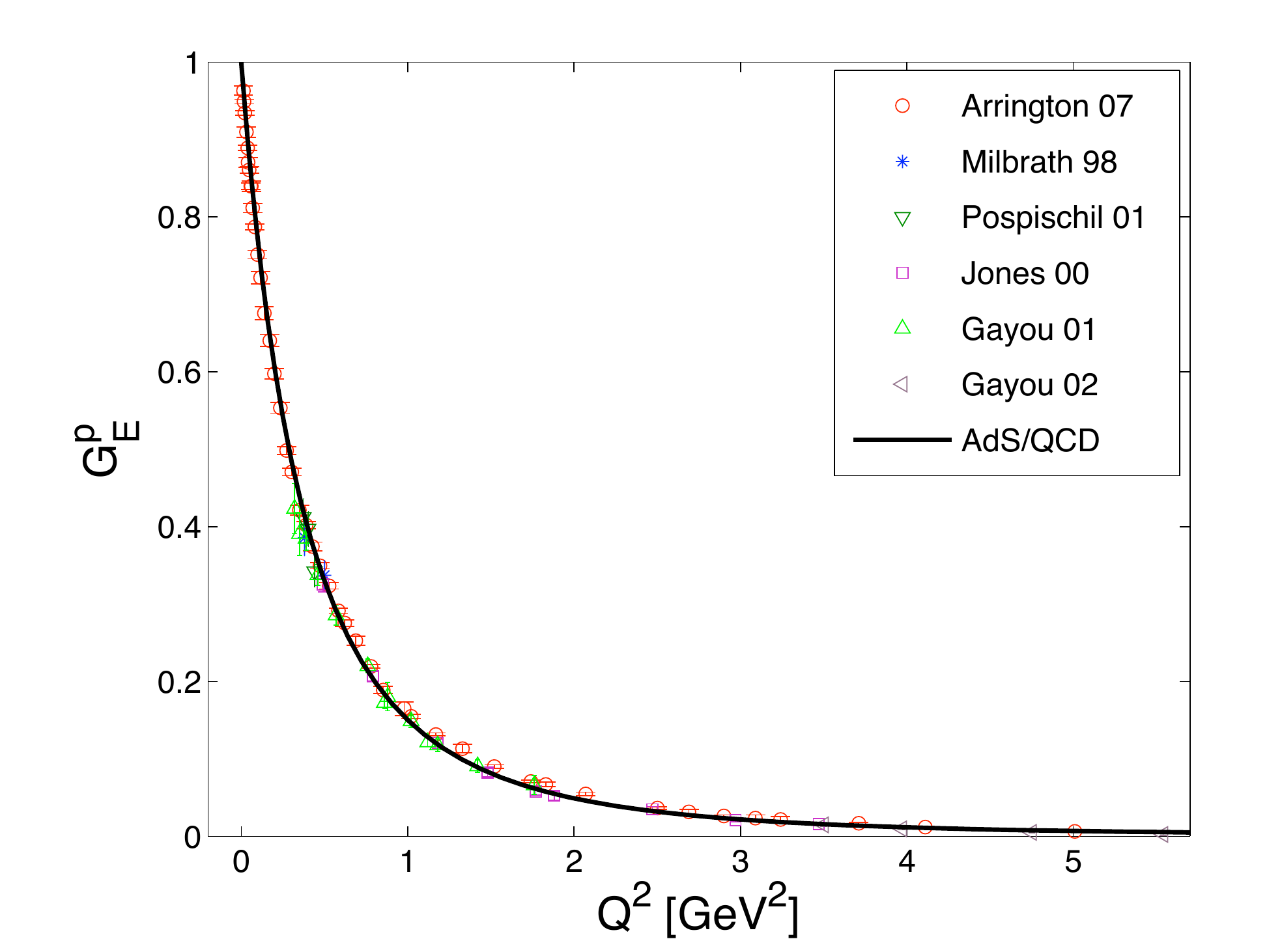}
\small{(b)}\includegraphics[width=7.5cm,height=7.5cm,clip]{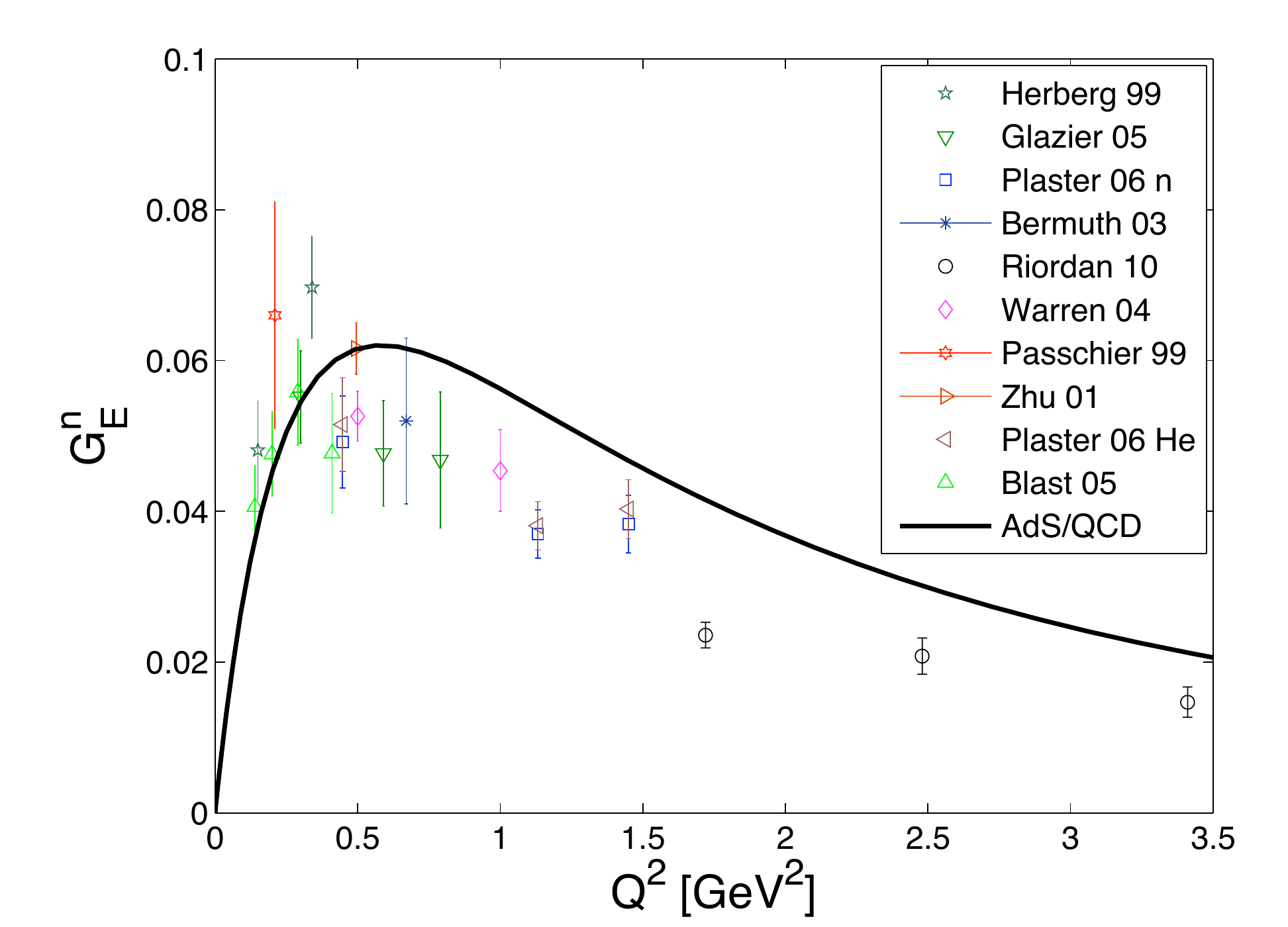}
\caption{\label{GpGn}(Color online)   (a) Sachs form factor $G_E(Q^2)$ for  proton. The experimental data are taken from Refs.\cite{Gay1,Gay2,Arr,Milbrath,Posp,Jones}, and (b)$G_E(Q^2)$ for  neutron. The experimental data are taken from Refs. \cite{Herb,Glaz,Plast,Berm,Riordan,Pass,Zhu,War,Blast}.}
\end{figure}
The Dirac and Pauli form factors for the nucleons are related to the  first moment of the valence GPDs  \cite{BDH,diehl}
\be
F_1^p(t) &=& \int_0^1 dx(\frac{2}{3} H_v^u(x,t)-\frac{1}{3}H_v^d(x,t)), \nonumber\\
F_1^n(t) &=& \int_0^1 dx(\frac{2}{3} H_v^d(x,t)-\frac{1}{3}H_v^u(x,t)), \nonumber\\
F_2^p(t) &=& \int_0^1 dx(\frac{2}{3} E_v^u(x,t)-\frac{1}{3}E_v^d(x,t)), \label{FF}\\
F_2^n(t) &=& \int_0^1 dx(\frac{2}{3} E_v^d(x,t)-\frac{1}{3}E_v^u(x,t)). \nonumber 
\ee
 Here $x$ is the fraction of the longitudinal momentum carried by the  quark and the GPDs for valence quark $q$ are  defined as $H_v^q(x,t)=H^q(x,0,t)+H^q(-x,0,t);$  $ E_v^q(x,t)=E^q(x,0,t)+E^q(-x,0,t). $
The bulk-to-boundary propagator, Eq. (\ref{propagator}), can be written in a simple integral form \cite{Rad,BT2}
\be
\!\! V(Q^2,z)=\kappa^2z^2\int_0^1\frac{dx}{(1-x)^2} x^{Q^2/(4\kappa^2)} e^{-\kappa^2 z^2 x/(1-x)}.
 \ee
We use the  integral form of the bulk-to-boundary propagator in the formulas for the form factors in AdS space to extract the GPDs using the formulas in Eq. (\ref{FF}).
The valence  GPDs are related to the flavor form factors  by the sum rules
\be
\int_0^1 dx H_v^q(x,t)&=&F_1^q(t),\\
\int_0^1 dx E_v^q(x,t)& = & F_2^q(t).
\ee
The GPDs in this model have been extensively studied in both momentum and impact parameter spaces in \cite{CM}.
We use these formulas to evaluate the flavor form factors from the GPDs and then compare them with the experimental results.
\begin{figure}[htb]
\small{(a)}\includegraphics[width=7cm,height=7cm,clip]{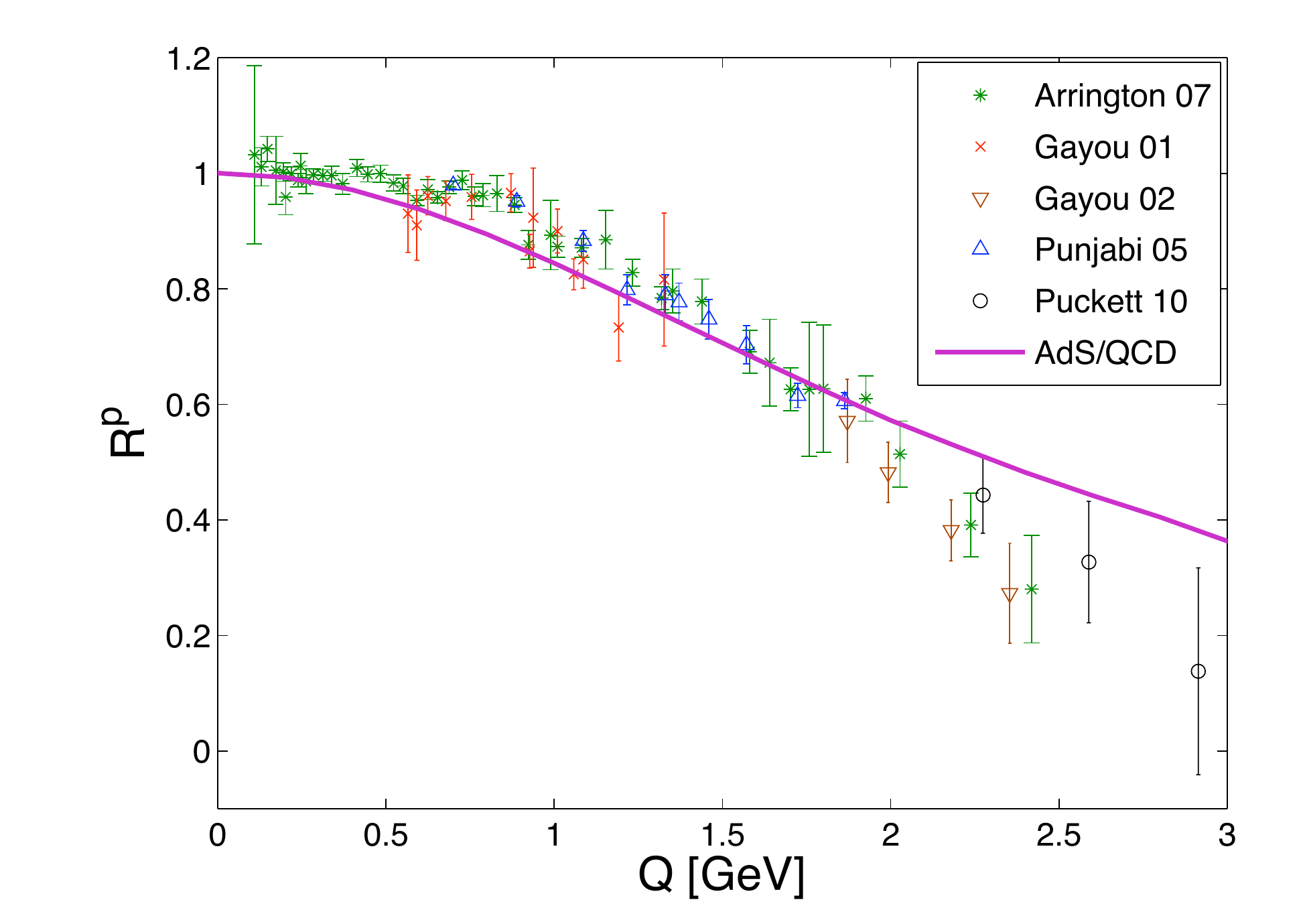}
\small{(b)}\includegraphics[width=7cm,height=7cm,clip]{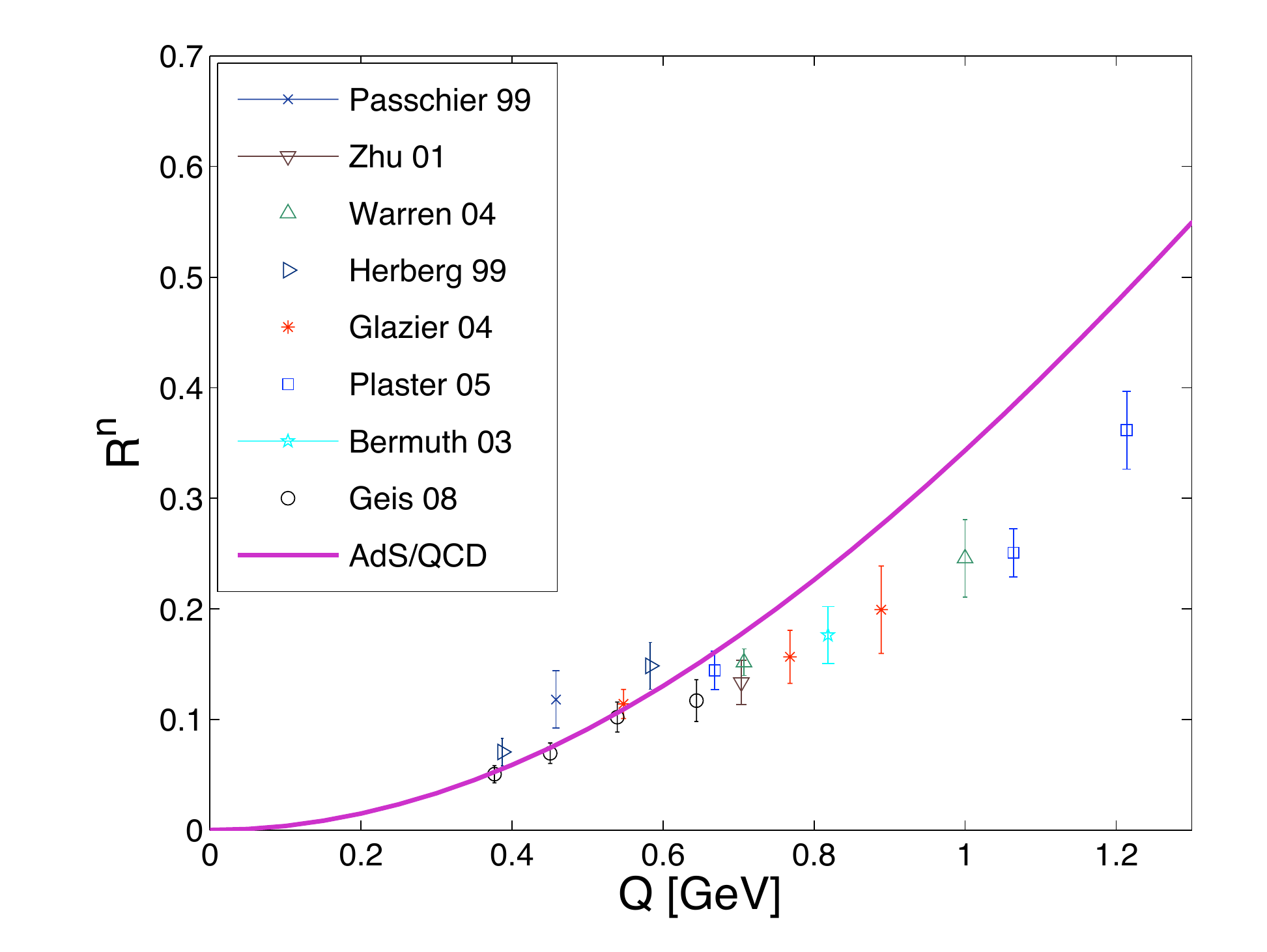}
\caption{\label{RpRn}(Color online)  (a) $R^p$ plotted against $Q=\sqrt{-t}$. Experimental data are taken from Refs. \cite{Gay1,Gay2,Arr,Pun,Puck}.   (b) $R^n$ plotted against $Q$. Experimental data are taken from Refs.\cite{Pass,Zhu,War,Herb,Glaz,Plast,Berm,Geis}.}
\end{figure}
\begin{figure}[htp]
\begin{minipage}[c]{0.98\textwidth}
\small{(a)}
\includegraphics[width=7cm,height=6cm,clip]{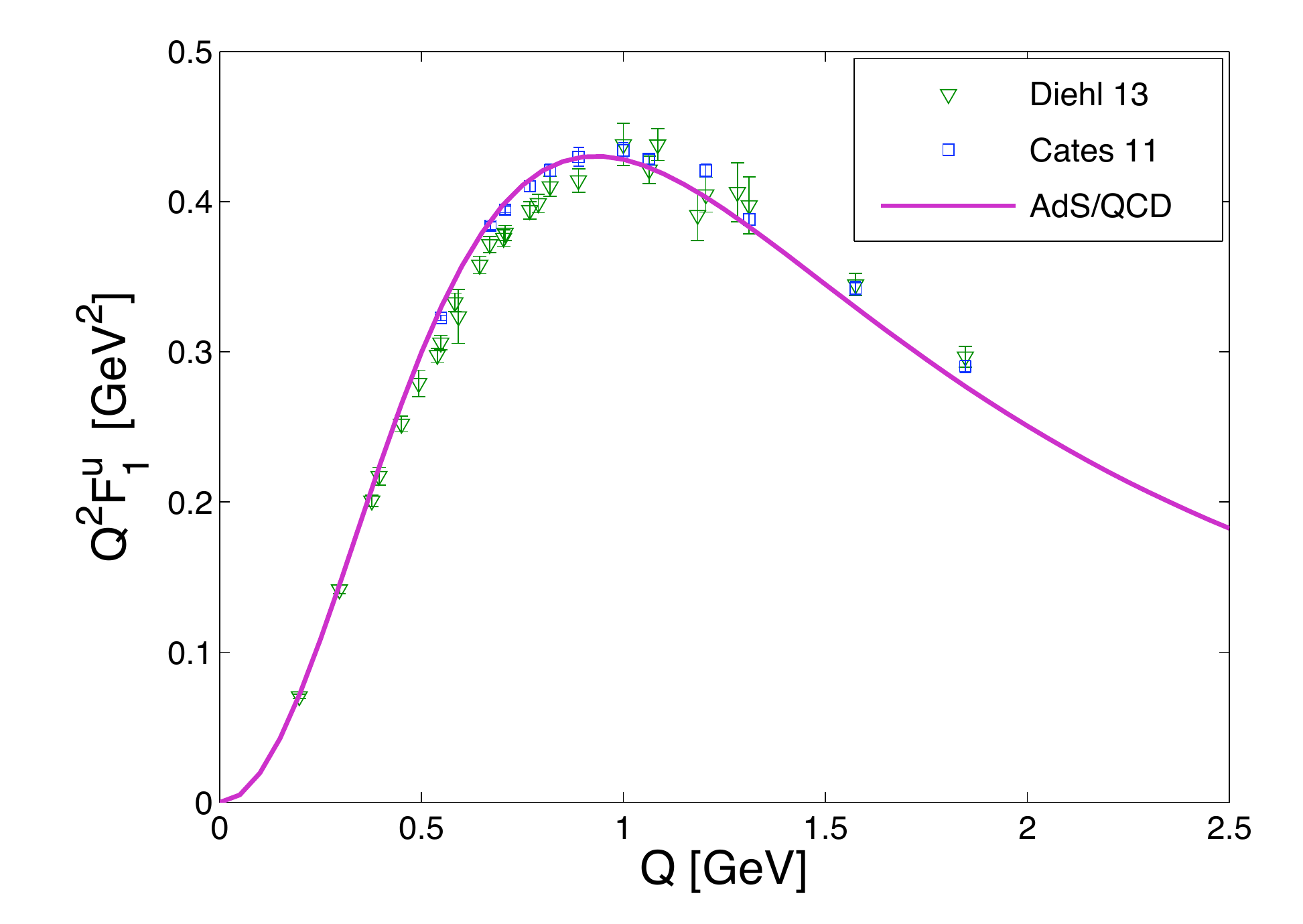}
\hspace{0.1cm}%
\small{(b)}\includegraphics[width=7cm,height=6cm,clip]{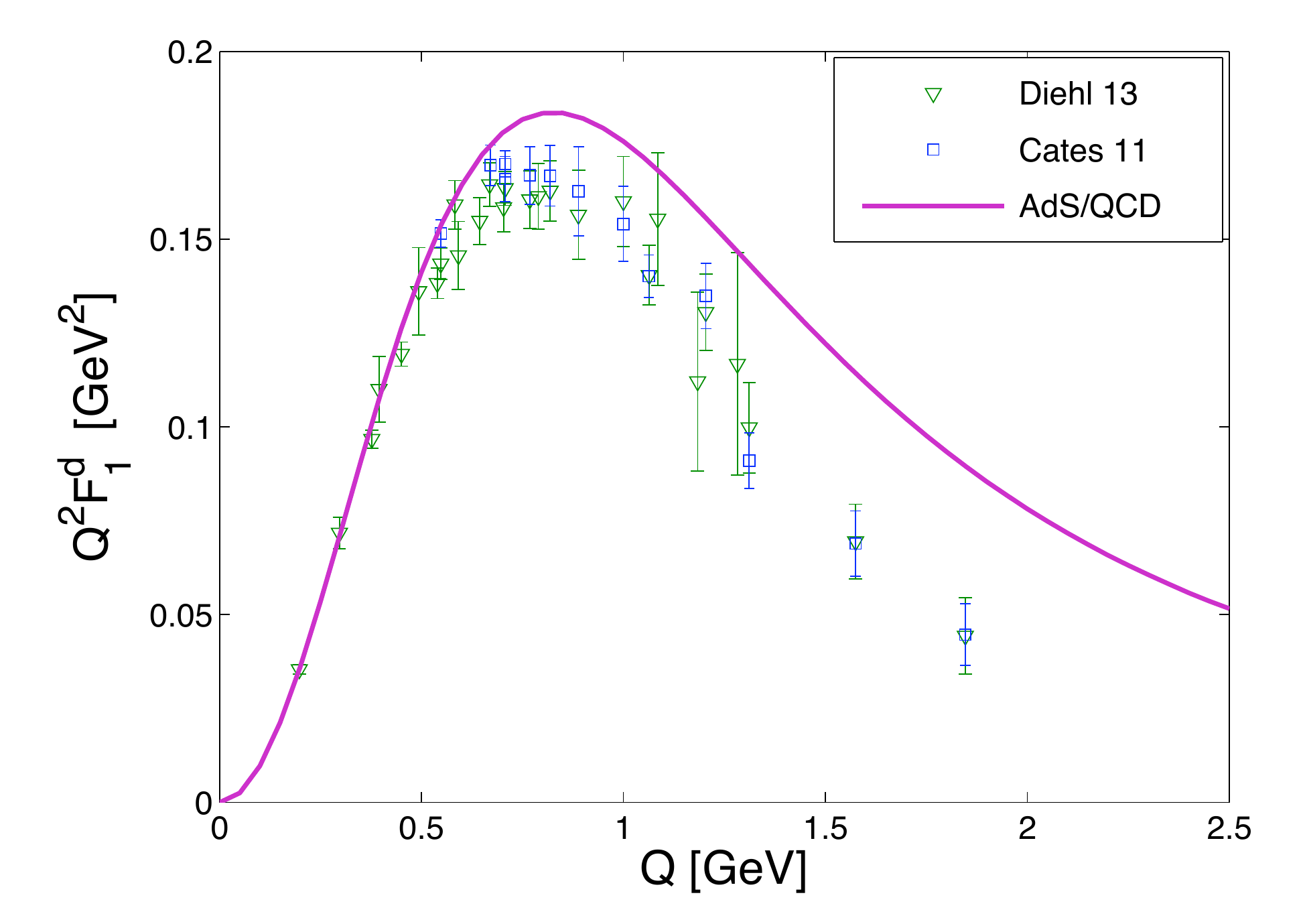}
\end{minipage}
\begin{minipage}[c]{0.98\textwidth}
\small{(c)}\includegraphics[width=7cm,height=6cm,clip]{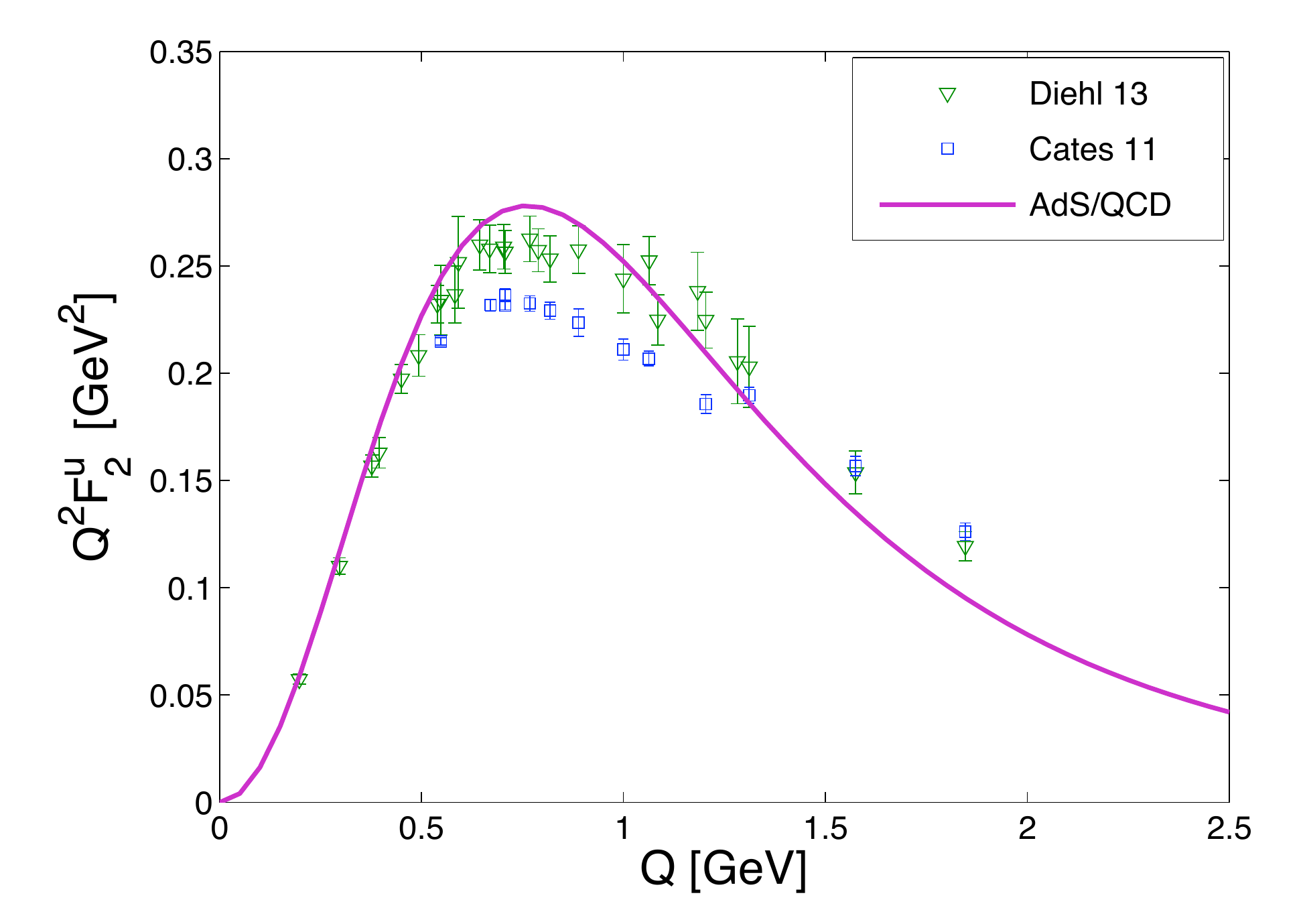}
\hspace{0.1cm}%
\small{(d)}\includegraphics[width=7cm,height=6cm,clip]{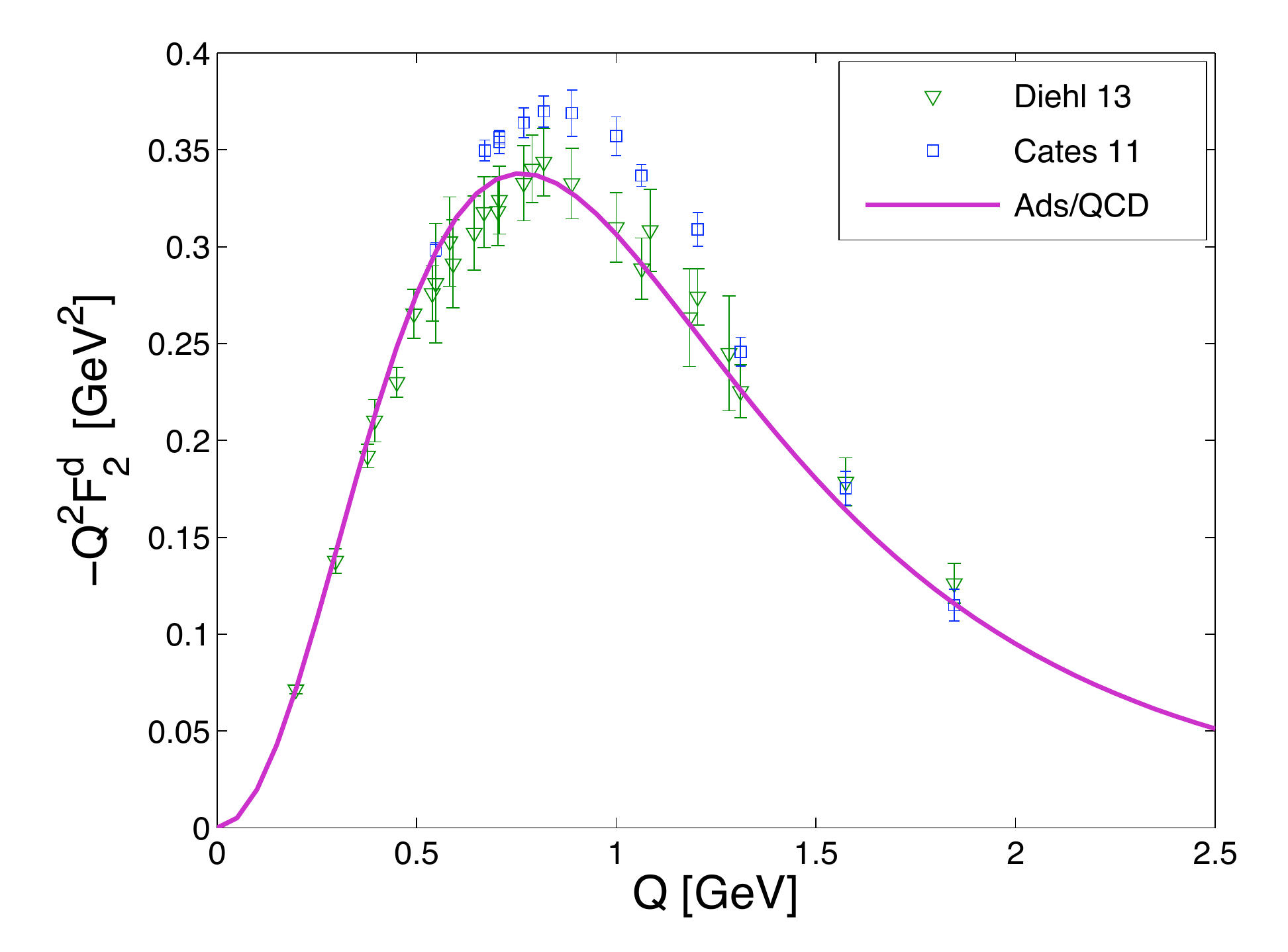}
\end{minipage}
\begin{minipage}[c]{0.98\textwidth}
\small{(e)}\includegraphics[width=7cm,height=6cm,clip]{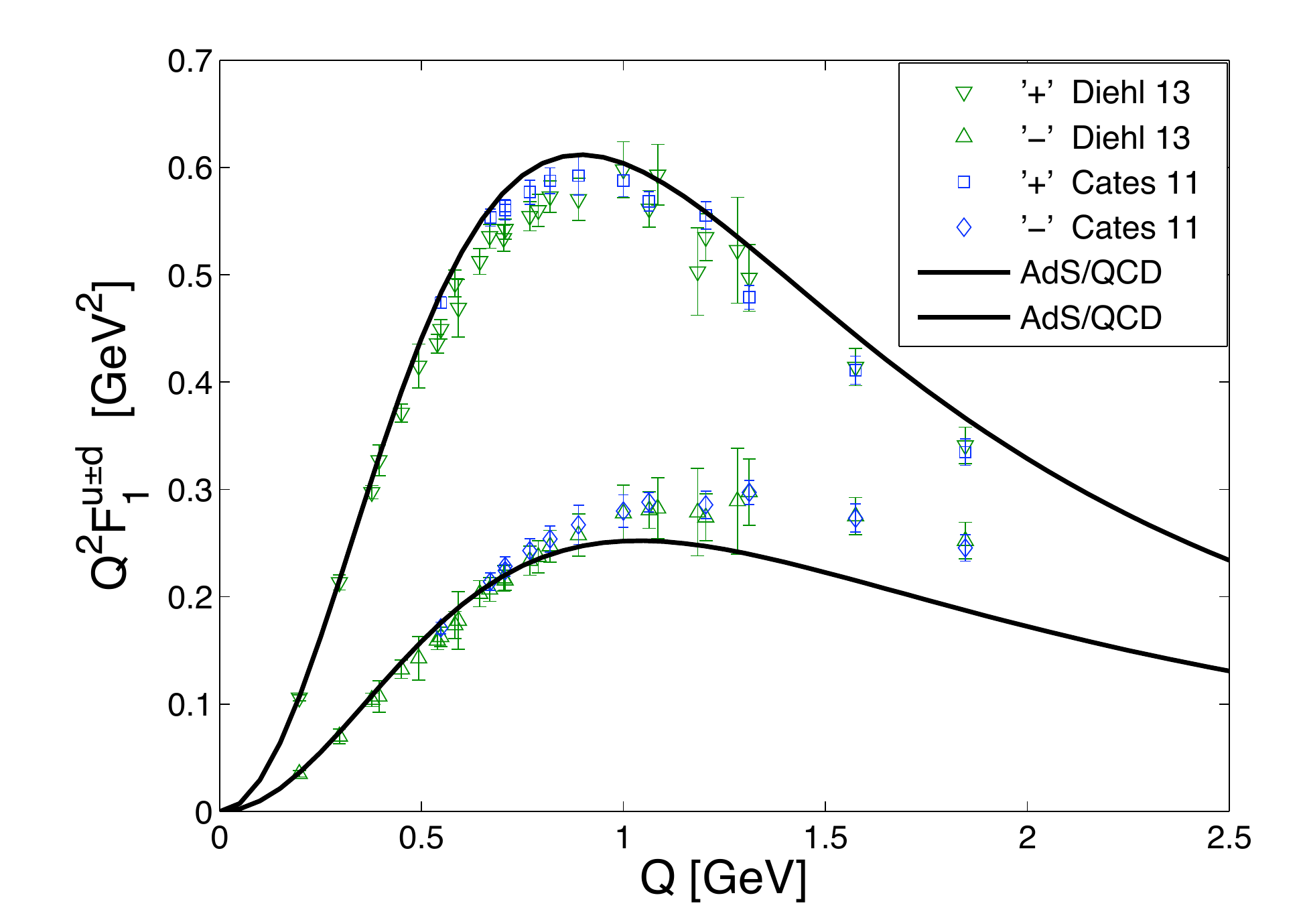}
\hspace{0.1cm}%
\small{(f)}\includegraphics[width=7cm,height=6cm,clip]{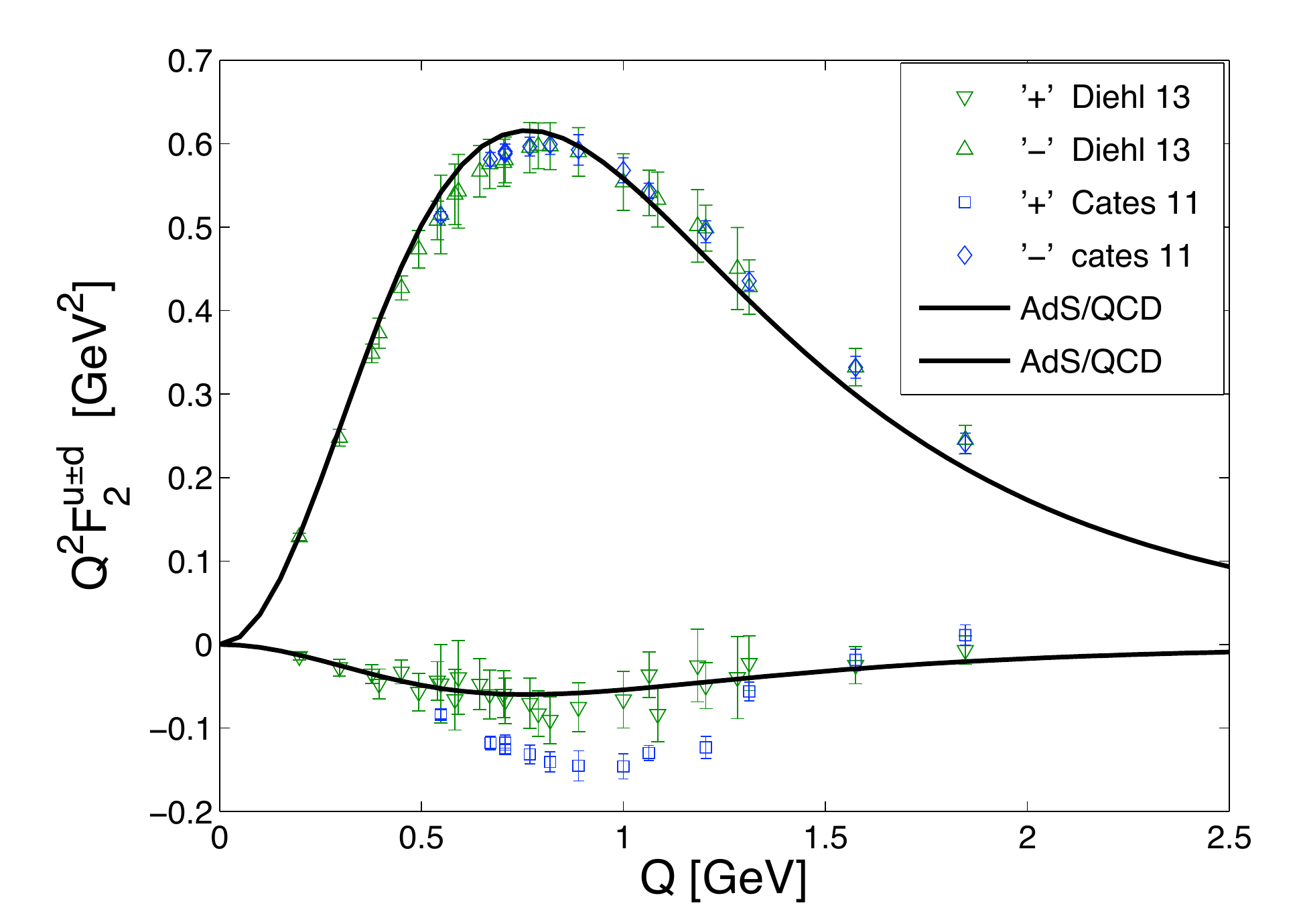}
\end{minipage}
\caption{\label{FF_flavors}(Color online) Plots of  flavor form factors for $u$ and $d$ quarks. In (e) and (f) the data legend $'+'$ stands for $F^{u+d}$ and  $'-'$ stands for $F^{u-d}$. $Q=\sqrt{-t}$. The experimental data are taken from \cite{Cates,diehl13}.}
\end{figure}

\section{Comparisons with experimental data}\label{comp}

In Fig.\ref{Fp_ratio}, we have shown the fit of our results with  experimental data of proton form factors. We get excellent agreement with the data for $\kappa=0. 4066$ GeV. After making the necessary subtraction of $-4\kappa^2$ as argued in \cite{BT2},  we get the nucleon mass corresponding to the above value of $\kappa$ as $M=0.813$ GeV. 
 All the plots for nucleon and favor form factors are done with this fixed value of $\kappa$. The Dirac and Pauli form factors for the nucleons have also been studied in \cite{BT2} where the parameter $\kappa$ was determined by using a different fitting procedure. Though their nucleon form factors are quite similar to ours, the flavor decompositions which provide us the information about the contributions of different quarks to the nucleon form factors agree better with experimental results in our method. 
  The nucleon form factors in AdS/QCD have  also been calculated using another model \cite{AC} and  with higher Fock states\cite{hf1}.  In all cases, the form factors for the neutron were found  not to agree with the experiments so well as those of proton. The  light front quark model results derived in this paper also show the similar behavior. For illustration,  we have shown the comparison of the electric form factor $G_E(Q^2)$ for both proton and neutron with the experimental data in Fig.\ref{GpGn}.
  
\begin{figure}[htbp]
\begin{minipage}[c]{0.98\textwidth}
\small{(a)}\includegraphics[width=7.5cm,height=7cm,clip]{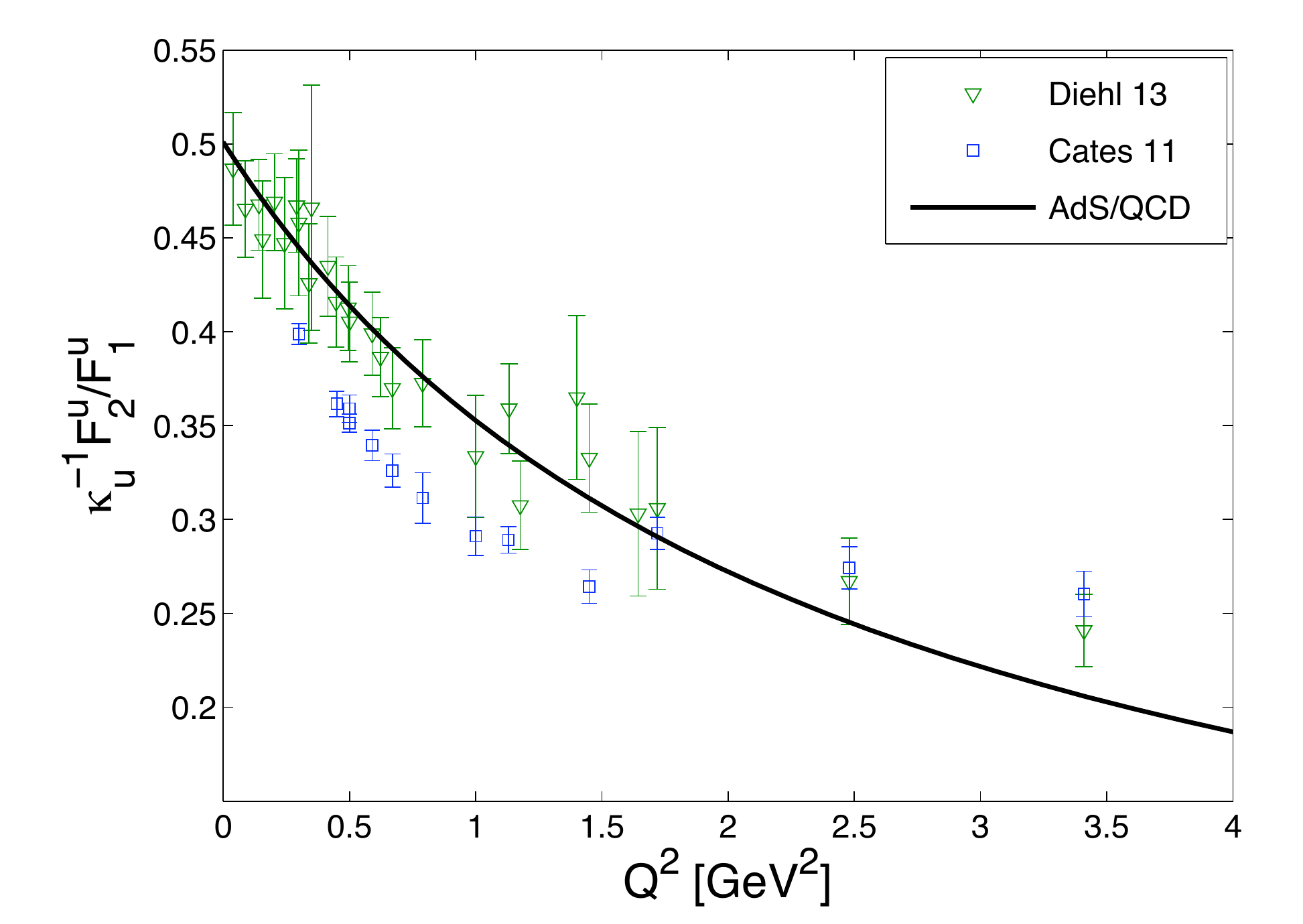}
\hspace{0.1cm}
\small{(b)}\includegraphics[width=7.5cm,height=7cm, clip]{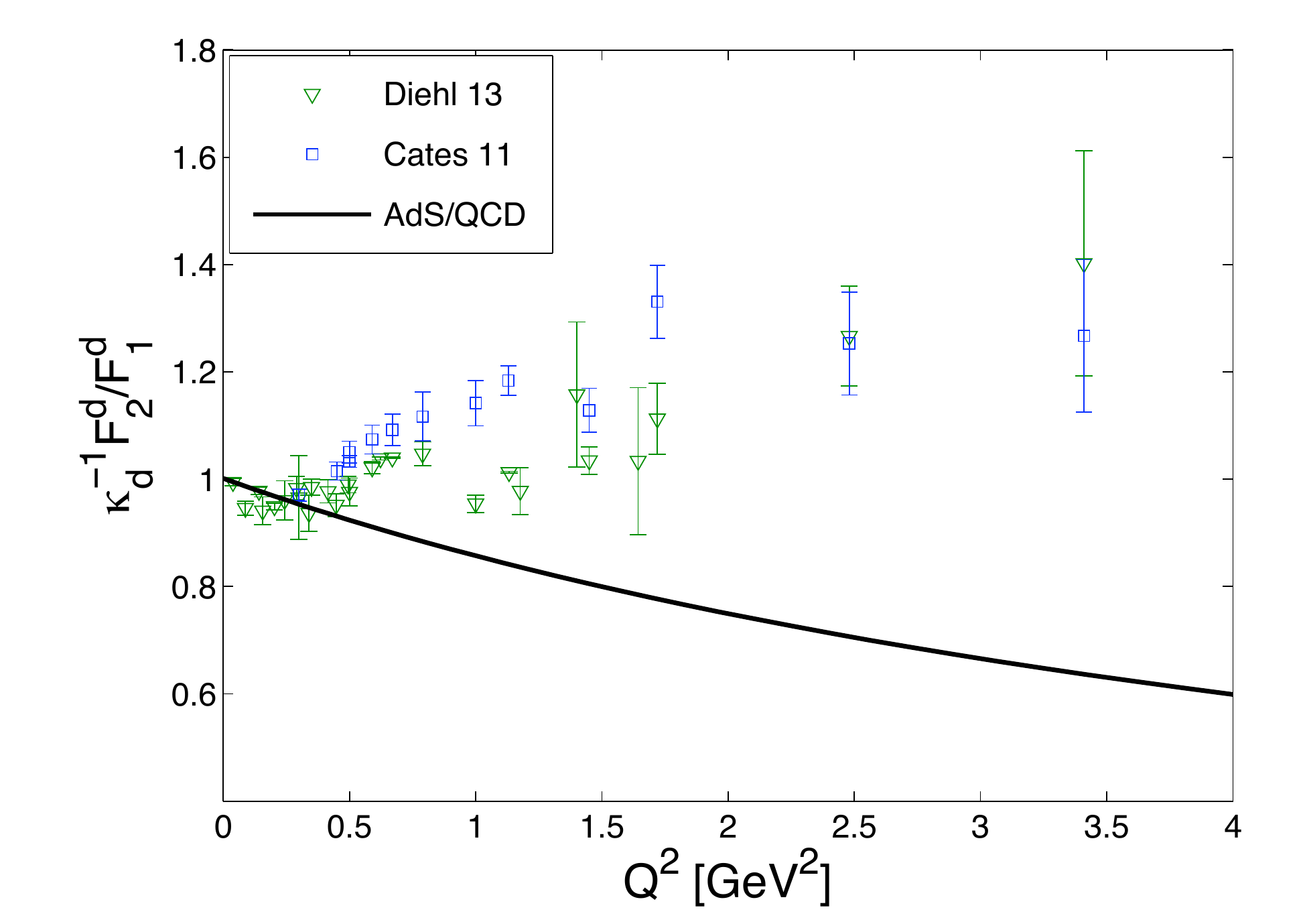}
\end{minipage}
\caption{\label{flavor_ratio}(Color online)   (a) $F_2^u/{(\kappa_u F_1^u)}$ plotted against $Q^2$.    (b) same as (a), but for $d$ quark. The AdS/QCD result for $d$-quark is similar to the RCQM result  as shown in Fig. 2 in Ref. \cite{Cates}. The experimental data are taken from \cite{Cates,diehl13}.}
\end{figure}

In Fig.\ref{RpRn} we have plotted the ratios $R^p$ and $R^n$ from data and the AdS/QCD results.  As remarked before,the agreement for $R^n$ is not so good, but considering that the AdS/QCD just gives a semiclassical approximation of the nucleons, the agreement is  actually not bad.  From the Sachs form factors we can also compute the electromagnetic radii of the nucleons. We quote the results here, the experimental  values quoted within the square brackets are taken from \cite{RPP}.
\be
\sqrt{\langle r^2_E\rangle_p}&=&0.8102  {\rm ~ fm},~~[ 0.877\pm0.005 {\rm ~fm}]; \nonumber\\
\sqrt{\langle r^2_M\rangle_p}&=&0.7826  {\rm ~ fm}, ~~[0.777\pm0.016 {\rm ~ fm}]; \nonumber\\
\langle r^2_E\rangle_n &=&-0.0882  {\rm ~ fm^2,}~~[ -0.1161\pm0.0022 {\rm ~fm}^2]; \nonumber\\
\sqrt{\langle r^2_M\rangle_n}&=&0.7965  {\rm ~ fm,}~~[ 0.862\pm0.009 {\rm ~ fm}].\nonumber
\ee
The electromagnetic radii for the  nucleons have previously been calculated in  AdS/QCD models in \cite{BT2,AC,vega}.

\begin{figure}[htbp]
\small{(a)}\includegraphics[width=7.5cm,height=7cm,clip]{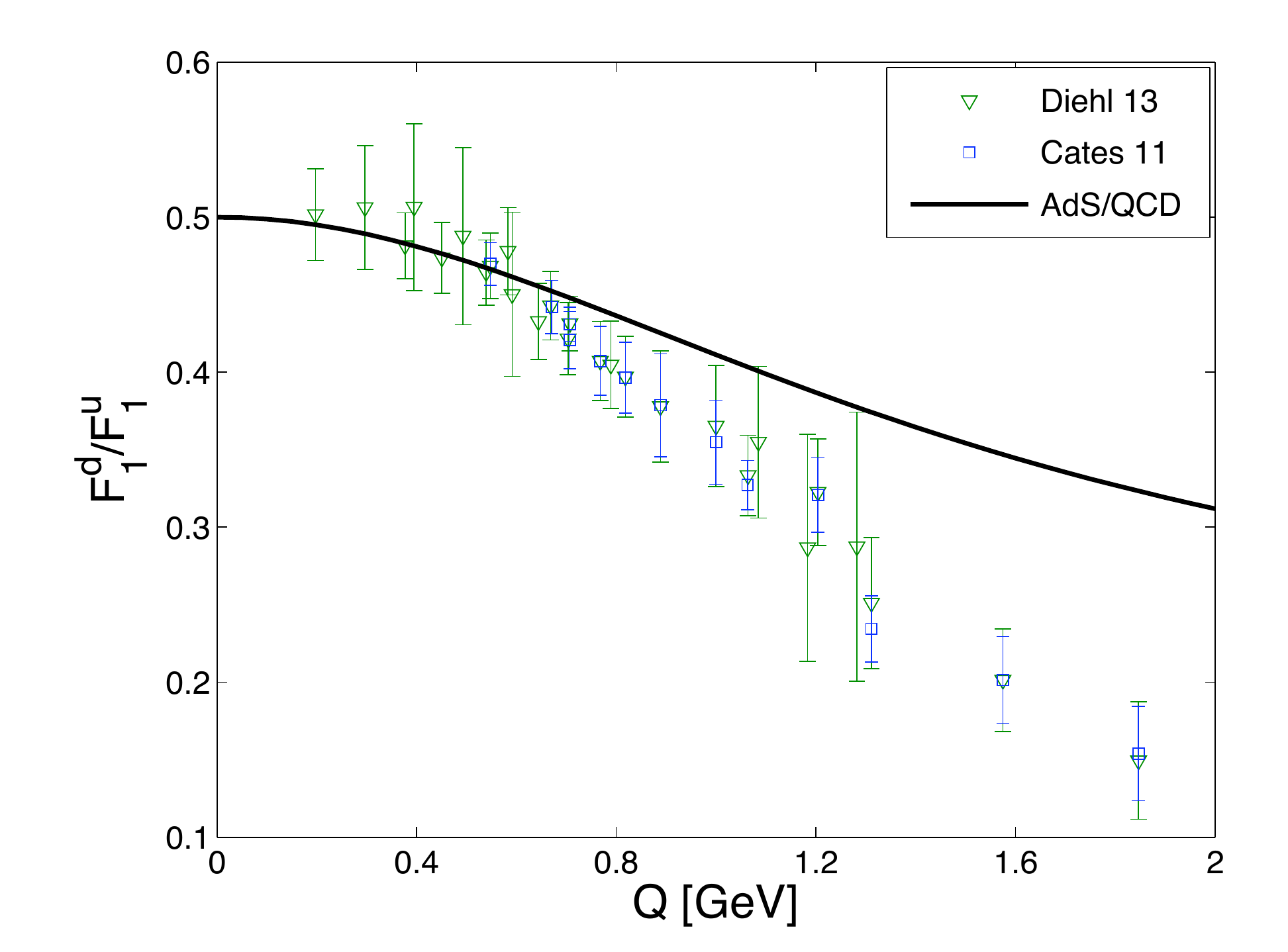}
\small{(b)}\includegraphics[width=7.5cm,height=7cm, clip]{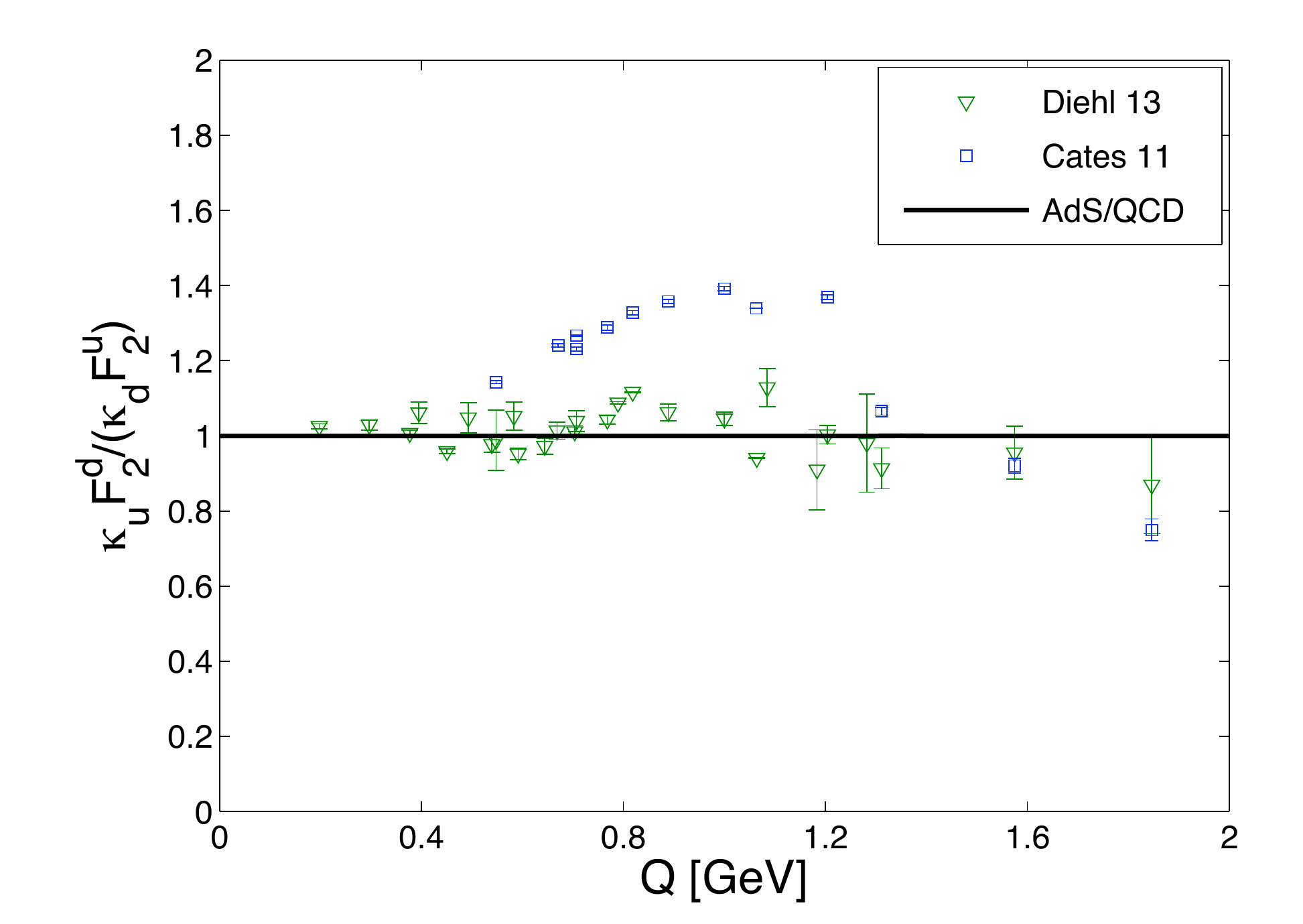}
\caption{\label{flavor_ratio2}(Color online)   Ratios of the flavor  form factors, (a) $ F_1^d/F_1^u$ and (b) $\kappa_uF_2^d/{\kappa_d F_2^u}$ plotted against $Q=\sqrt{-t}$. The experimental data are taken from \cite{Cates,diehl13}.}
\end{figure}

Now from GPDs, we calculate the flavor form factors for $u$ and $d$ quarks and compare with the experimental data.   In \cite{Cates}, the flavor form factors are extracted from the polarization data of the ratios of nucleon Sachs form factors.  Diehl and  Kroll \cite{diehl13} have recently  extracted the flavor form factors from  $R^p, R^n, G_M^{p/n}/G^{p/n}_{dipole}$. Since the data points for proton and neutron Sachs form factors  are in general not at same $t$ values, they used an interpolation method. For the comparison with our results for flavor form factors, we have shown both the data. It is clear from Fig.\ref{FF_flavors}, that the AdS/QCD results for $u$-quark form factors are in excellent agreement with the data while $F_1^d(t)$ does not agree so well with the data. The experimental data show that at large $Q^2$, both the $d$-quark form factors $F_1^d$ and $F_2^d$ fall off faster than the corresponding $u$-quark form factors.  For $F_1^d$, at small $Q^2$, the AdS/QCD results are in agreement with the data but the deviation increases at larger $Q^2$.  It is important  to note here that  other models also fail to reproduce the form factors data for $d$ quark\cite{Qattan}. But, from Fig.\ref{FF_flavors}(d), we can see that the AdS/QCD reproduces  $F_2^d$ data extremely well.
In Fig.\ref{flavor_ratio}, we have shown the ratios of flavor form factors for each flavor.  Again our results for   $u$-quark agree with the data and  deviate for $d$-quark. The AdS/QCD result for $d$-quark is similar to the RCQM result\cite{Cates}. In the last figure, Fig.\ref{flavor_ratio2}, we have shown the ratios $F_1^d/F_1^u$ and $\kappa_uF_2^d/(\kappa_d F_2^u)$. Since $F_1^d$ deviates from data for large $Q^2$, the ratio $F_1^d/F_1^u$ deviates from the data at large $Q^2$.  The ratio $\kappa_uF_2^d/(\kappa_d F_2^u)$ in AdS/QCD is   constant for all $Q^2$ values and is 1.0 where the experimental data are clustered around. 
\begin{figure}[htbp]
\begin{minipage}[c]{0.95\textwidth}
\small{(a)}
\includegraphics[width=7cm,height=6cm,clip]{Gep_vs_q2_k0575.pdf}
\hspace{0.1cm}%
\small{(b)}\includegraphics[width=7cm,height=6cm,clip]{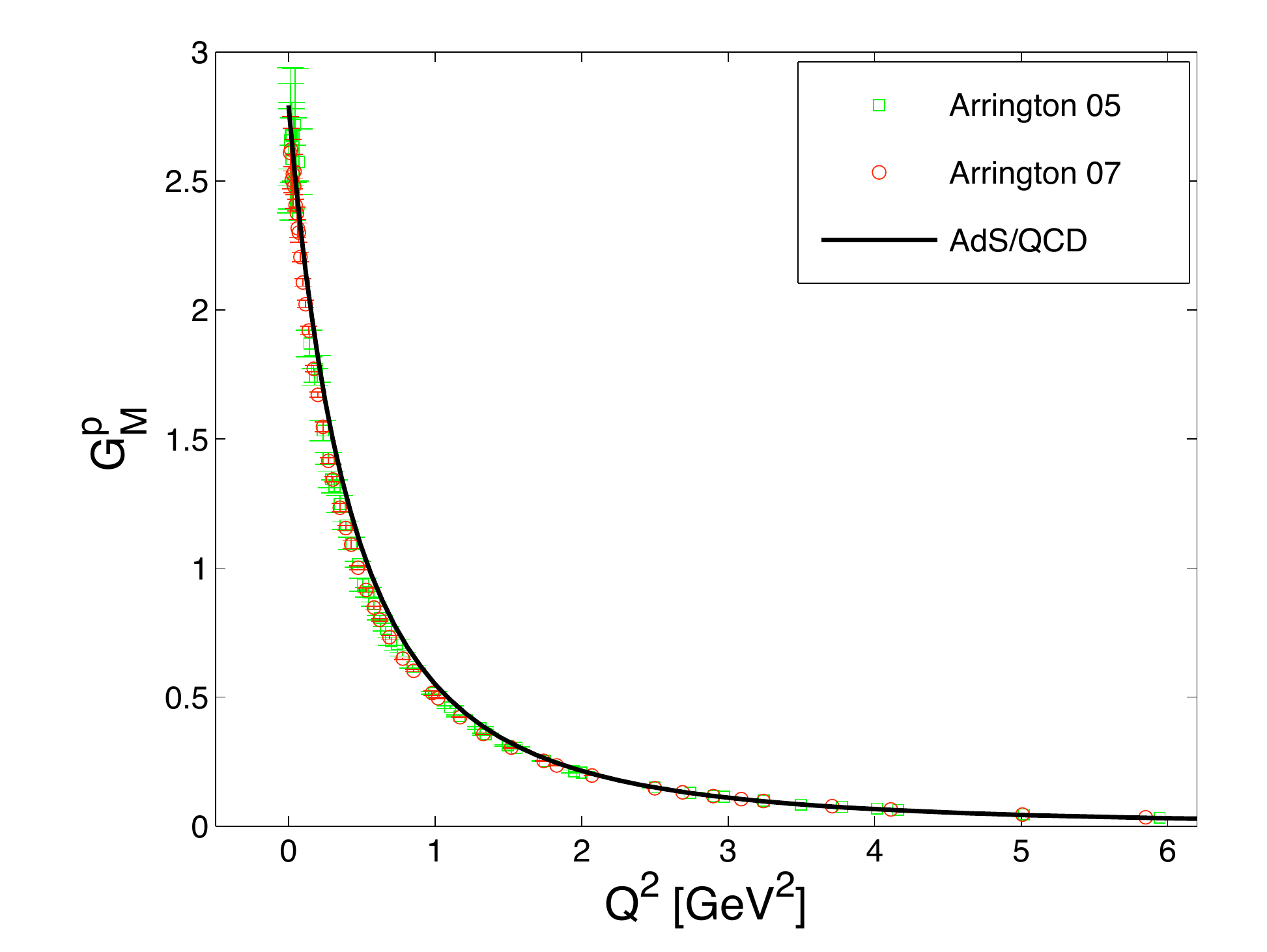}
\end{minipage}
\begin{minipage}[c]{0.98\textwidth}
\small{(c)}\includegraphics[width=7cm,height=6cm,clip]{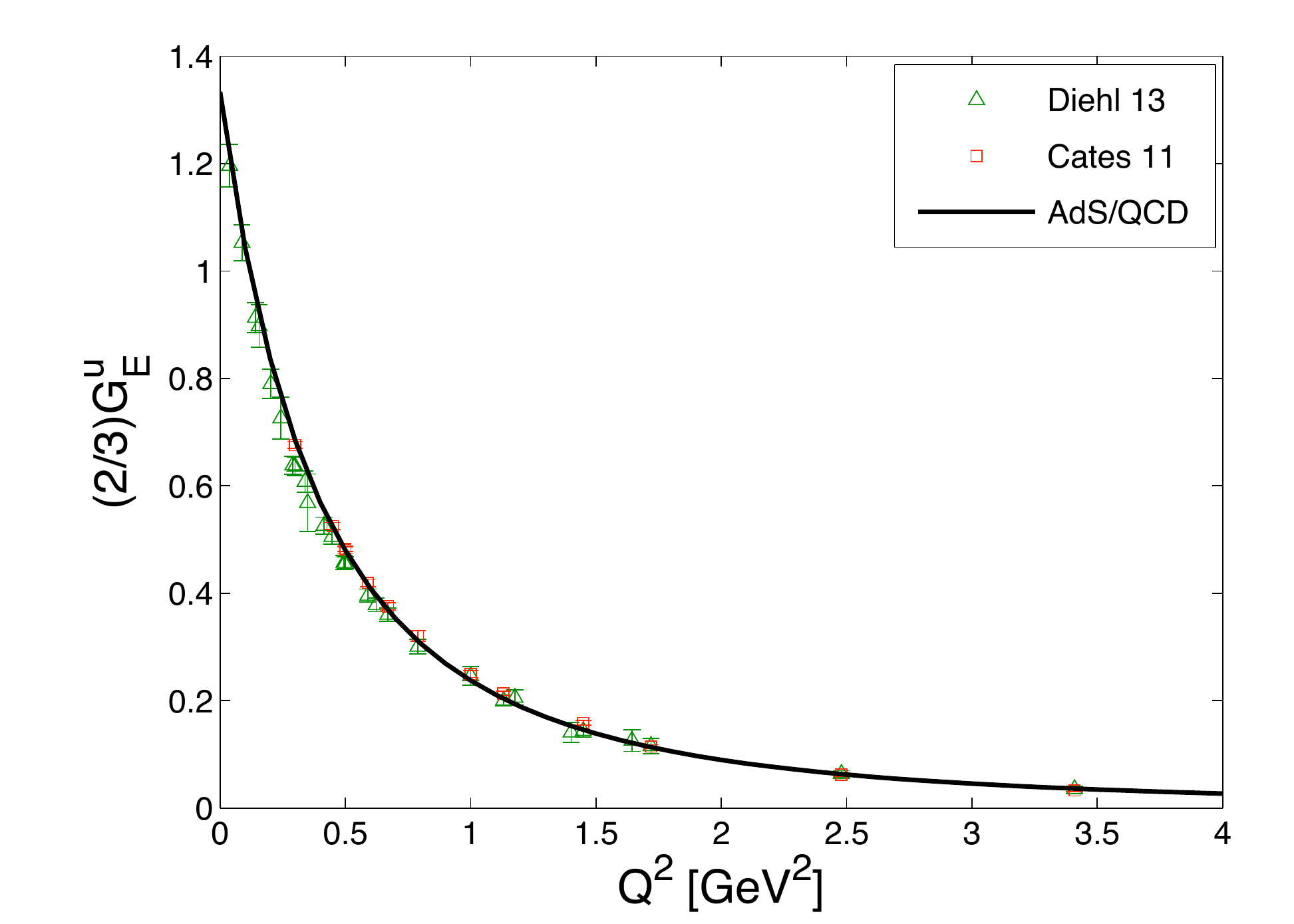}
\hspace{0.1cm}%
\small{(d)}\includegraphics[width=7cm,height=6cm,clip]{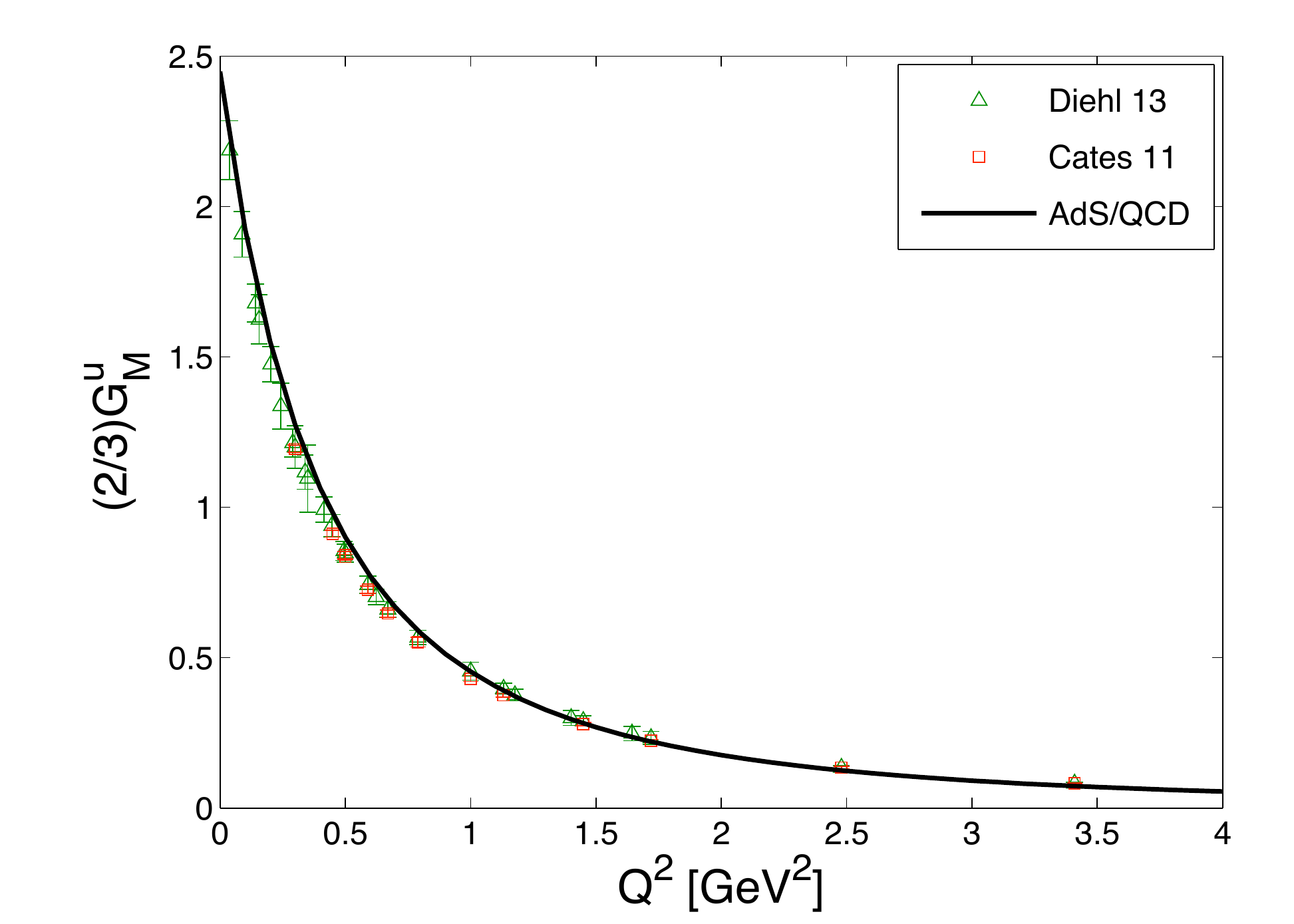}
\end{minipage}
\begin{minipage}[c]{0.98\textwidth}
\small{(e)}\includegraphics[width=7cm,height=6cm,clip]{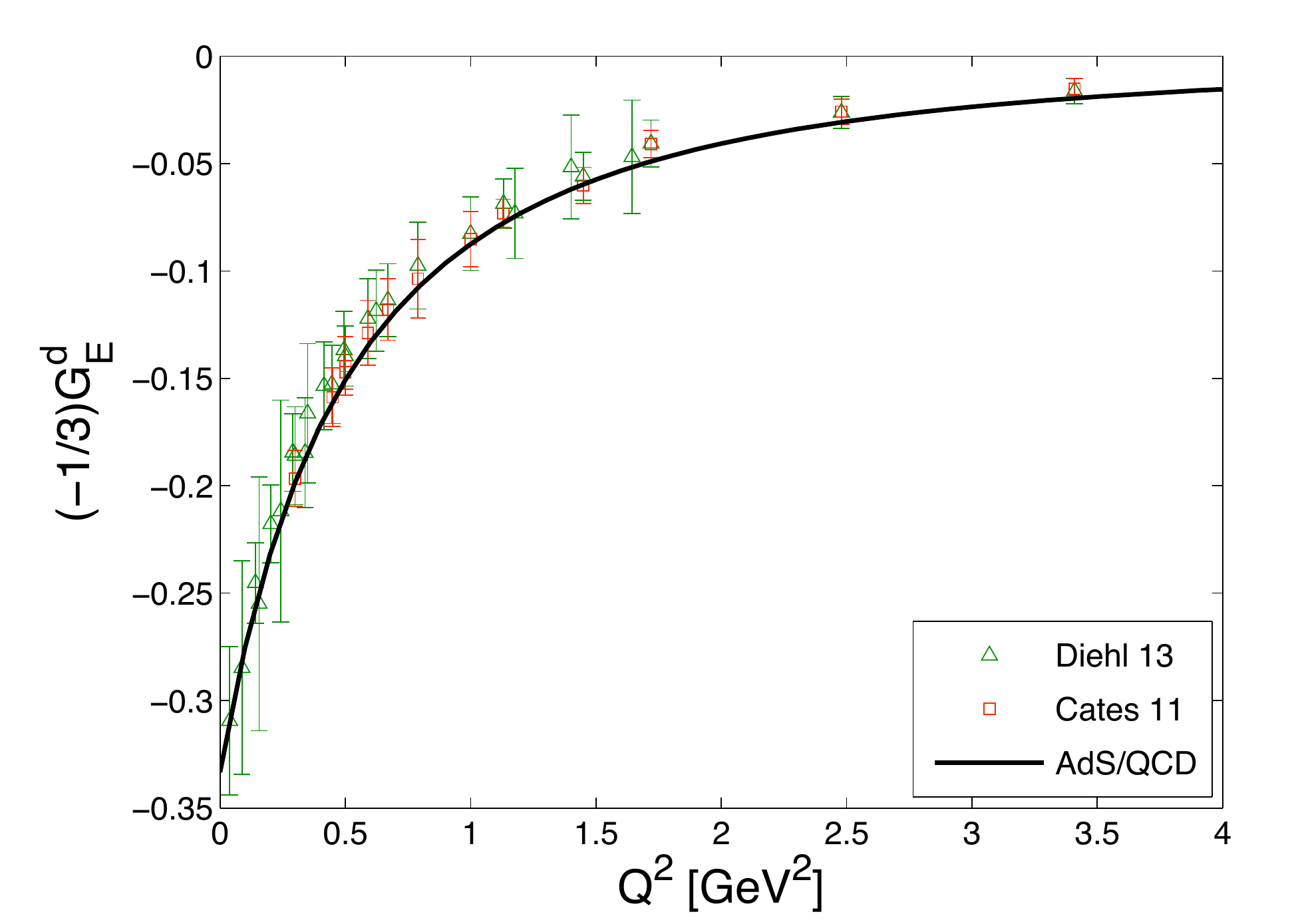}
\hspace{0.1cm}%
\small{(f)}\includegraphics[width=7cm,height=6cm,clip]{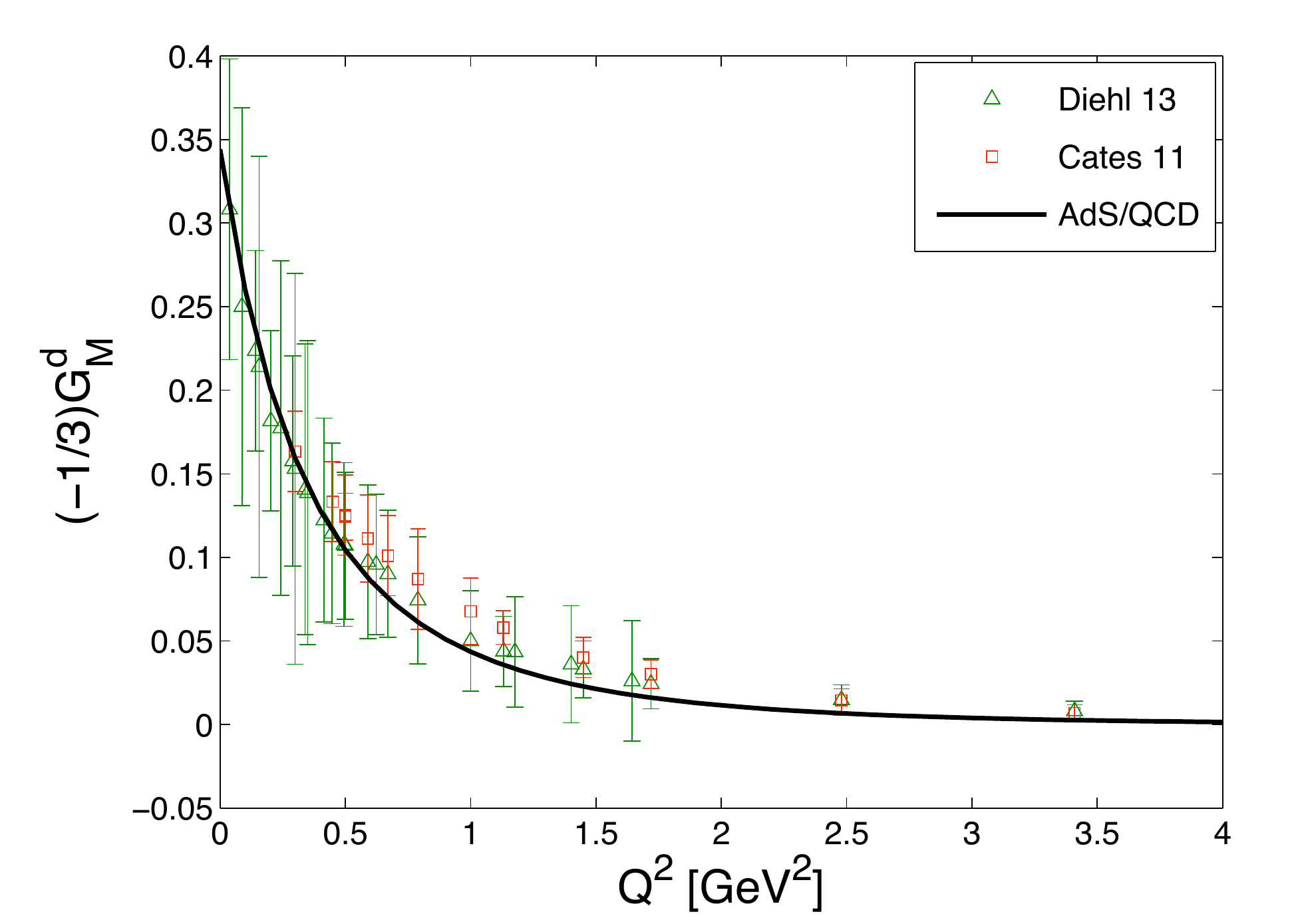}
\end{minipage}
\caption{\label{Gp_flavors}(Color online) Plots of  flavor decomposition of the Sachs form factors for the proton. (a) and (b) represent  the Sachs  form factors for the proton, (the experimental data are taken from \cite{Herb,Glaz,Plast,Berm,Riordan,War,Pass,Zhu,Blast} and \cite{Arr,Arr2} ; (c)-(f) represent  the contributions from different flavors.
The experimental data are taken from \cite{Cates,diehl13}.}
\end{figure}
\begin{figure}[htbp]
\begin{minipage}[c]{0.98\textwidth}
\small{(a)}
\includegraphics[width=7cm,height=6cm,clip]{Gen_vs_q2_k0575.pdf}
\hspace{0.1cm}%
\small{(b)}\includegraphics[width=7cm,height=6cm,clip]{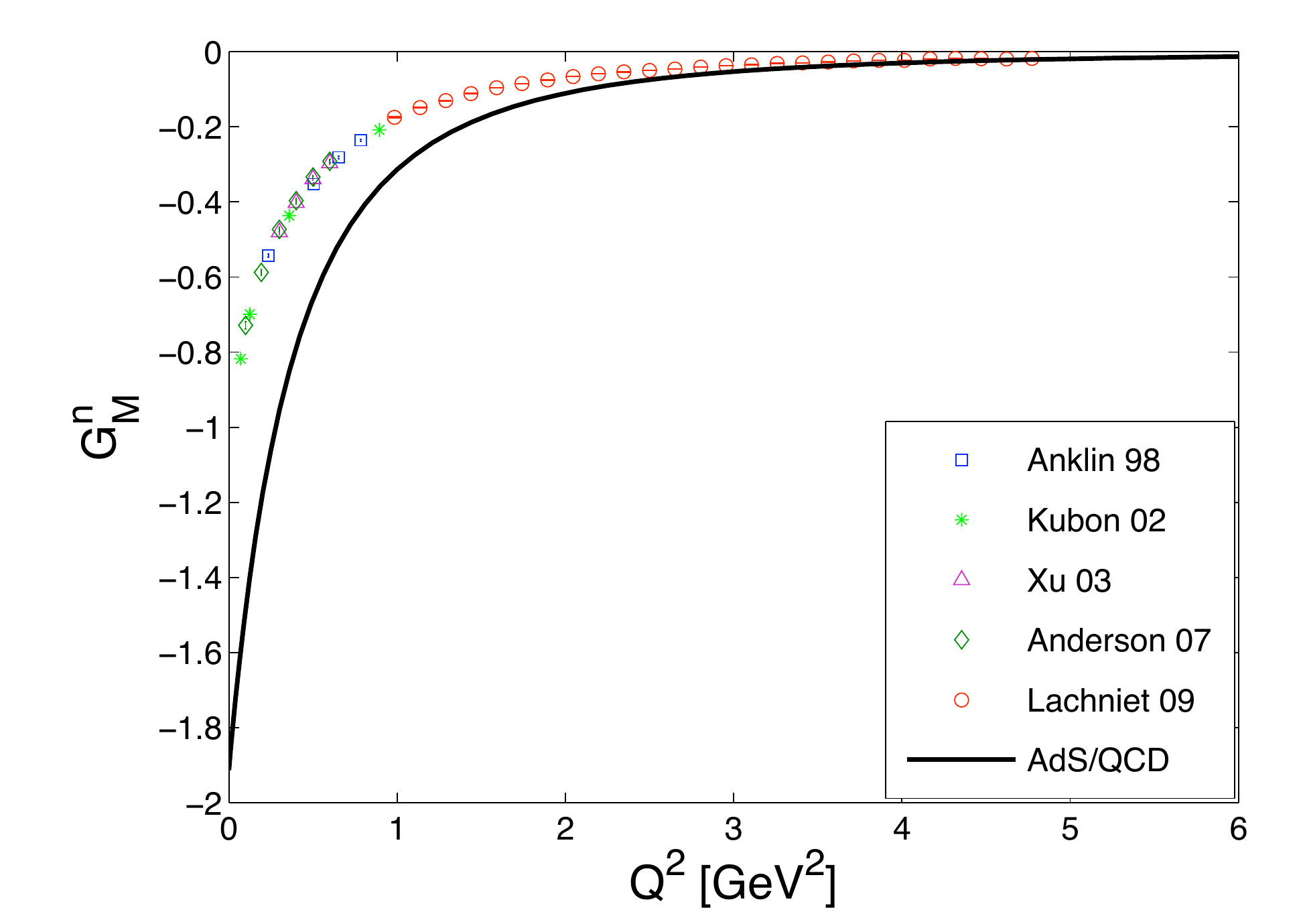}
\end{minipage}
\begin{minipage}[c]{0.98\textwidth}
\small{(c)}\includegraphics[width=7cm,height=6cm,clip]{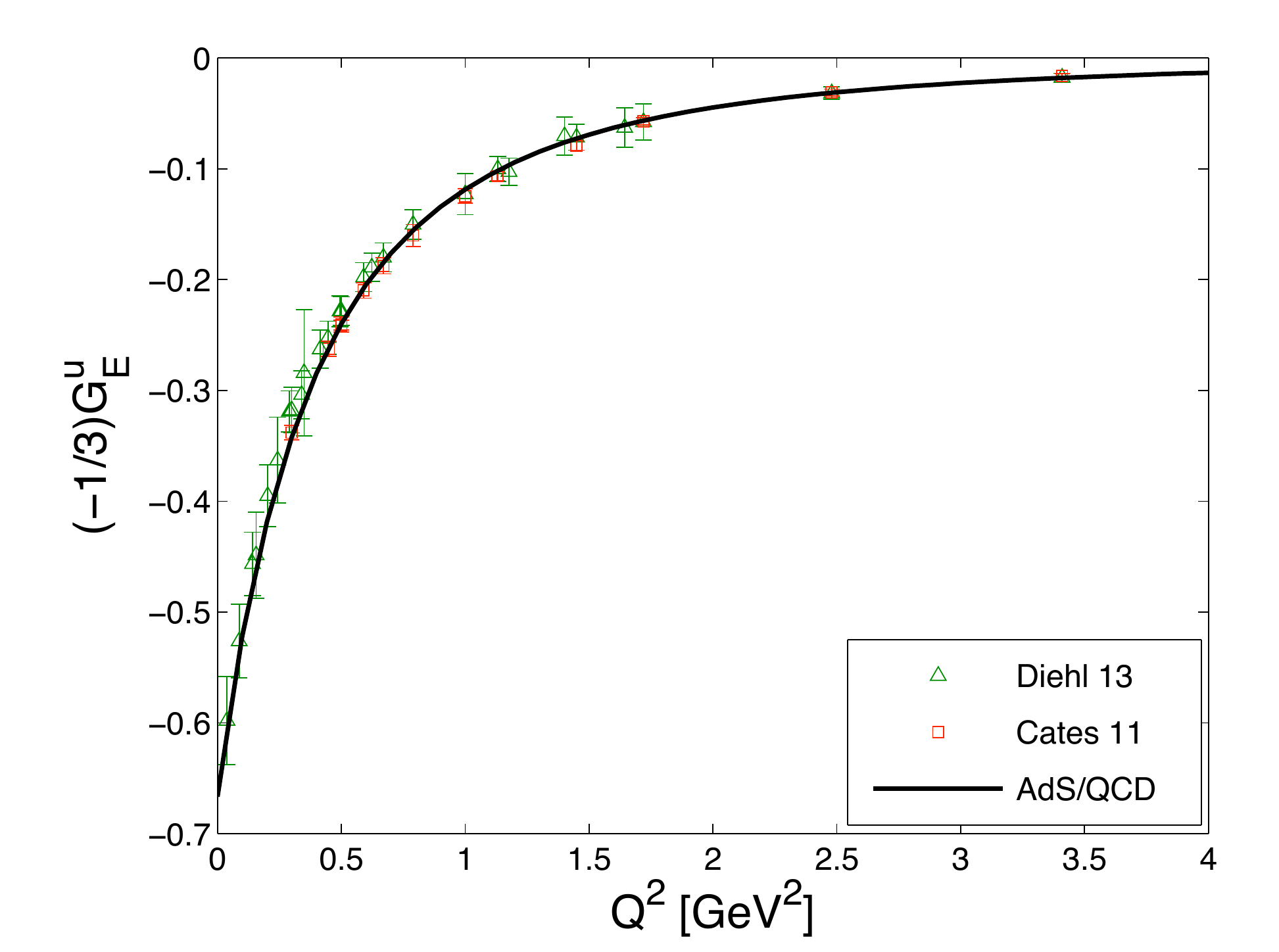}
\hspace{0.1cm}%
\small{(d)}\includegraphics[width=7cm,height=6cm,clip]{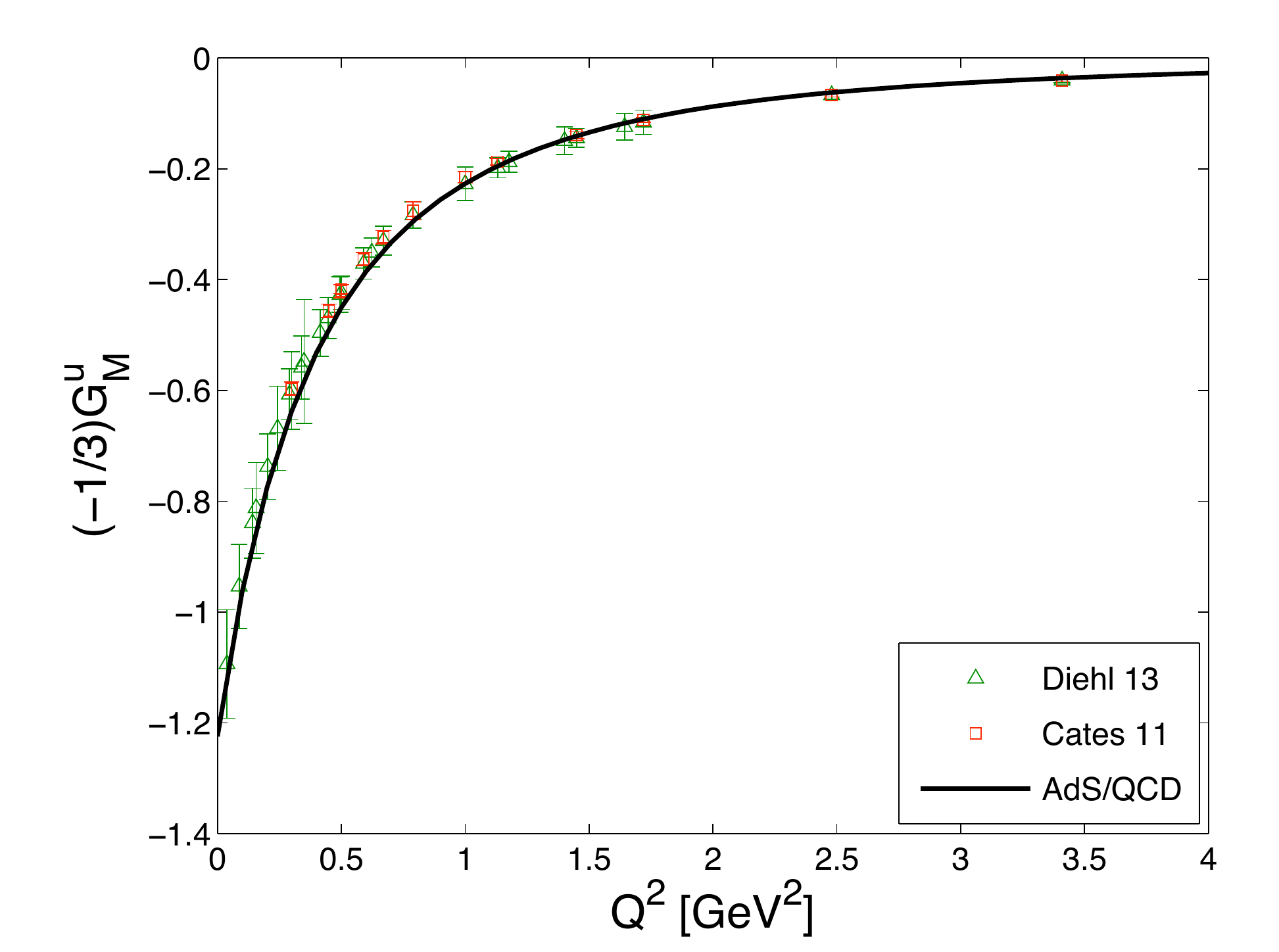}
\end{minipage}
\begin{minipage}[c]{0.98\textwidth}
\small{(e)}\includegraphics[width=7cm,height=6cm,clip]{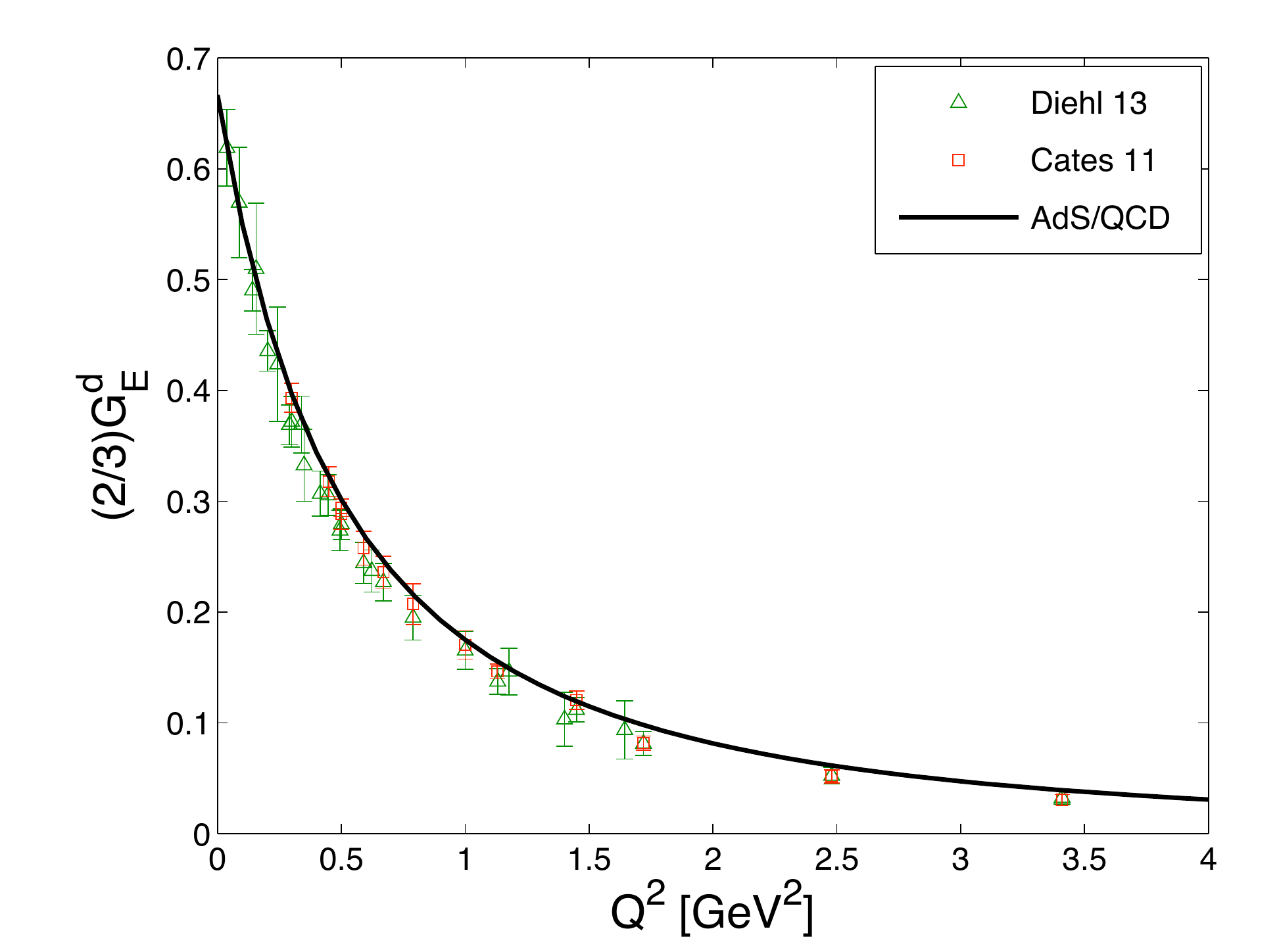}
\hspace{0.1cm}%
\small{(f)}\includegraphics[width=7cm,height=6cm,clip]{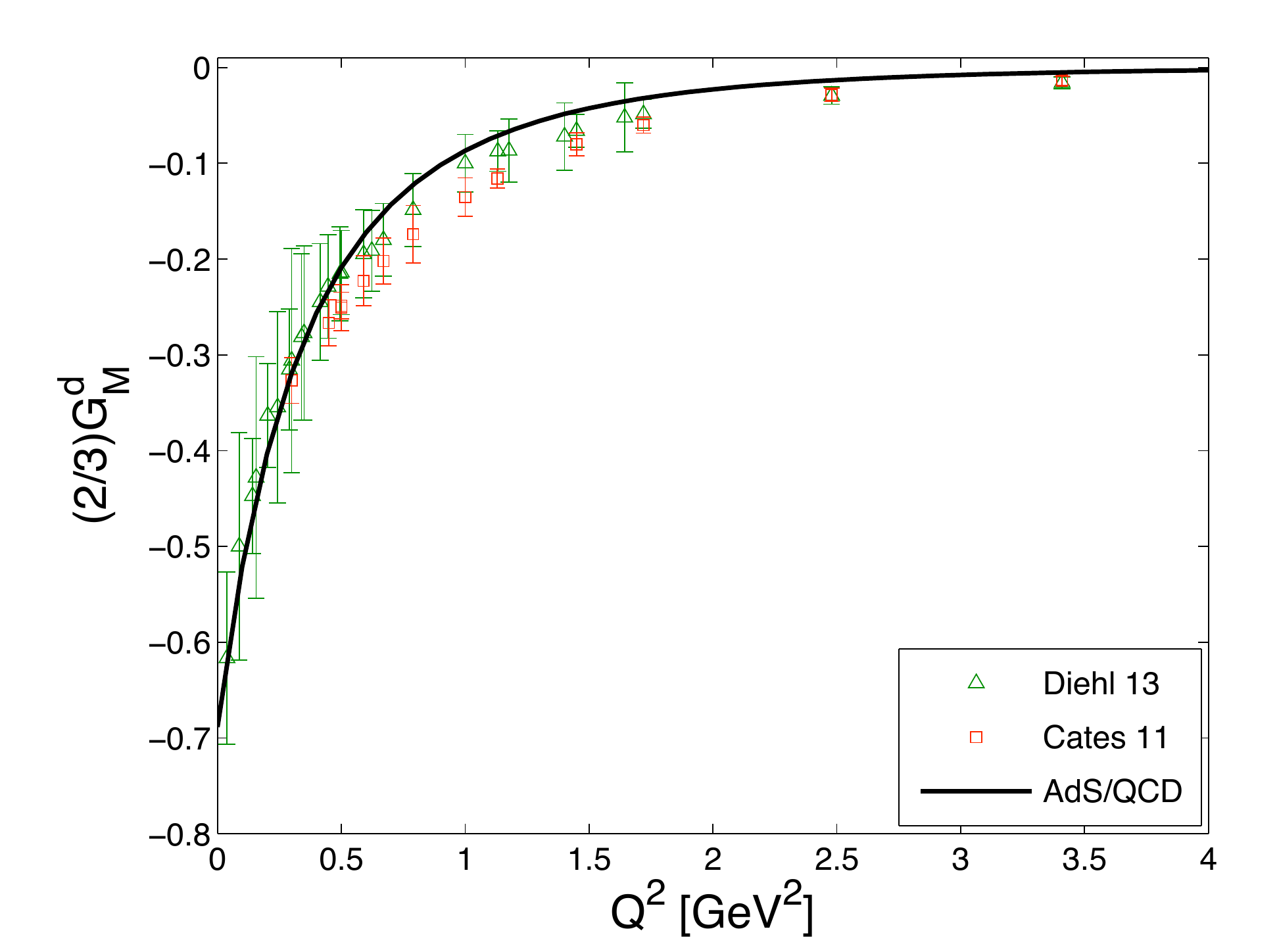}
\end{minipage}
\caption{\label{Gn_flavors}(Color online) Plots of  flavor decomposition of the Sachs form factors for the neutron. (a) and (b) represent  the Sachs  form factors for the neutron,( the experimental data are taken from \cite{Arr,Milbrath,Posp,Jones,Gay1,Gay2} and \cite{Anklin,Kubon,Xu,Anderson,Lachniet}));  (c)-(f)  represent contributions from different flavors.
The experimental data are taken from \cite{Cates,diehl13}.}
\end{figure}

In Fig. \ref{Gp_flavors}(a) and (b)  we have shown the Sachs electromagnetic form factors $G_E^p$ and $G_M^p$ for the proton. The  flavor contributions coming to these form factors $e_u G_{E/M}^u$ and $e_d G_{E/M}^d$ are shown 
in Fig. \ref{Gp_flavors}(c)-(f). Similarly the Sachs  form factors for the neutron and the flavor contributions $e_d G_{E/M}^u$ and $e_u G_{E/M}^d$ are shown in Fig.(\ref{Gn_flavors}). For the Sachs electromagnetic form factors of the proton,  the major contributions come from the $u$ quark,  the $d$ quark contributions is comparatively small. On the other hand, for the neutron,  both $u$ and $d$ quark contributions are comparable and are always of the same order of magnitude. For $G_E^n$, the up and down contributions are almost the  same but opposite in sign, and for $G_M^n$ both up and down quark contributions are of the same sign but the up quark contribution is  slightly stronger than the down quark.

\section{Summary and Conclusions}\label{concl}
 In this paper, we have calculated the nucleon form factors  and the contributions from individual flavors in a quark model in AdS/QCD.  The model assumes an SU(6) spin-flavor symmetry. The only parameter in the model is fixed by fitting to the data for the proton form factor ratio $Q^2 F_2^p/F_1^p$.  All other form factors are then calculated with the same value of the parameter.  The nucleon form factors and their ratios  agree well with the experimental data. We have also studied the flavor decompositions of the nucleon form factors for  $u$ and $d$ quarks. This is done by first extracting the GPDs for the $u$ and $d$ quarks from the Dirac and Pauli form factors. Since the first moments of the GPDs give the electromagnetic form factors, the flavor form factors are calculated from the moments of the  quark GPDs  in the model.
The results are compared with the available experimental data. The AdS/QCD results are found to be  in good agreement with the experimental data.  The AdS/QCD results for  the down quark  Dirac form factor $F_1^d(Q^2)$ do not agree with experimental data so well. The deviation increases for higher $Q^2$.
 But, it should be noted that this is very common with most of the models studied so far to evaluate the flavor form factors.  For the up quark, AdS/QCD results are in excellent agreement with the data.  From the results presented in this paper, we can infer that the deviations of the nucleon form factors from the experimental data can be attributed to the fact that the  down  flavor form factor $F_1^d$ does not have the correct behavior in this model.  We have also shown  that the Sachs electromagnetic form factors for the nucleons and their flavor decompositions in the model predicts the experimental results quite well.    We have also evaluated the electric and magnetic radii of the nucleons. They are   in agreement with the experimental values. 
 


\end{document}